\documentclass[a4paper, 11pt]{article}
\usepackage{amsmath,amsthm,amssymb,mathrsfs,mathtools,graphicx,color,yhmath}
\usepackage[titletoc,title]{appendix}

\usepackage{bm}
\usepackage{here}
\usepackage{bbm}
\usepackage[hang,small,bf]{caption}
\usepackage[subrefformat=parens]{subcaption}
\newcommand{\nc}{\newcommand}
\nc{\rnc}{\renewcommand}
\nc{\nn}{\nonumber}
\nc{\ms}{\mathsf}

\nc{\del}{{\partial}}
\rnc{\Im}{{\textrm{Im}\,}}
\rnc{\Re}{{\textrm{Re}\,}}

\nc{\bra}{\langle}
\nc{\ket}{\rangle}
\nc{\dr}{\mathrm{dr}}
\nc{\for}{\textrm{for}}
\nc{\s}{\sigma}
\nc{\la}{\lambda}
\nc{\g}{\boldsymbol{\mathcal{G}}}
\nc{\n}{\tilde{n}}

\nc{\A}{\textrm{A}}
\nc{\B}{\textrm{B}}
\nc{\C}{\textrm{C}}
\nc{\D}{\textrm{D}}
\nc{\tcr}{\textcolor{red}}
\nc{\tcb}{\textcolor{blue}}

\DeclareMathOperator{\sh}{sh}
\DeclareMathOperator{\ch}{ch}

\theoremstyle{definition}

\numberwithin{equation}{section}

\begin{document}

\title{Infinite temperature spin dc conductivity of the spin-1/2 XXZ chain }

\author{
Shinya Ae \thanks{E-mail: 1221701@ed.tus.ac.jp} 
\\\\
\textit{Department of Physics,
Tokyo University of Science,}\\
 \textit{Kagurazaka 1-3, Shinjuku-ku, Tokyo 162-8601, Japan} \\
\\\\
\\
}

\date{\today}

\maketitle

\begin{abstract}
Using the Bethe ansatz method and the TBA equations for the higher spin integrable XXZ chain, the regular zero frequency contribution to the spin current correlation 
(spin dc conductivity) is analyzed for the spin-1/2 XXZ chain with an anisotropy $0 \le \Delta <1$.
In the high temperature limit, we write down the dressed scattering kernels by one quasi-particle bare energies, which allows the exact evaluation of the infinite temperature spin dc conductivity $\mathcal{L}$.  
We find that $\mathcal{L}$ is discontinuous at all rational numbers of the anisotropy parameter $p_0=\pi/\cos^{-1}\Delta$ in the region $p_0 \ge 2$ with the gap increasing larger than the second power of growing magnetization on one quasi-particle. 
The isotropic $\Delta=1$ point is exceptional. Close to this point, $\mathcal{L}$ slowly increases in proportion to the first power of the magnetization.
On the other hand $\mathcal{L}$ is proportional to the second power of the magnetization when $p_0$ approaches irrational numbers.  
\end{abstract}

\section{Introduction and Summary}\label{Intro. and Summ.}
The Hamiltonian of the spin-1/2 XXZ model is given as follows for a chain of $L$ sites with periodic boundary conditions $\boldsymbol{S}_{L+1} \equiv \boldsymbol{S}_1$: 
\begin{equation}
H=J\sum_{i=1}^{L}\left(S^x_iS^x_{i+1}+S^y_iS^y_{i+1} +\Delta S^z_iS^z_{i+1}
\right)-2h\sum_{i=1}^L S^z_i, 
\label{Hamiltonian}\end{equation} 
where $S^{x,y,z}_i:=\s^{x,y,z}_i/2$ are Pauli's spin operators at site $i$, $J$ is the coupling constant, $h$ is the applied magnetic field and $\Delta$ is the anisotropy. The region $0 \le \Delta <1$ is parametrized by 
\begin{equation}
\Delta=\cos\theta, \quad 0< \theta=\frac{\pi}{p_0} \le \frac{\pi}{2}.
\label{anisotropy}\end{equation} 
The spin dc conductivity $\mathcal{L}(\beta)$ at inverse temperature $\beta=1/T \; (\textup{we set} \; k_B=1)$ is defined in parallel with the spin Drude weight $D(\beta)$ \cite{Klumper}. First, one defines the spin current density $j_i$ by the discrete continuity equation:
\begin{equation}
\partial _tS^z_i=-i[S^z_i,H]=-(j_i-j_{i-1}),
\end{equation}    
from which one obtains
\begin{equation}
j_i=i\frac{J}{2}(S^+_iS^-_{i+1}-S^-_iS^+_{i+1}); \quad S^\pm_i=S^x_i \pm iS^y_i
\end{equation}    
and the total spin current operator $J_0=\sum_i j_i$. Second, one can consider the spin conductivity $\s(\omega)$, a function of frequency $\omega$ based on the Kubo formula:
\begin{equation}
\sigma(\omega)
=\frac{i}{\omega}
\left(\frac{\bra H_\textup{kin} \ket}{L}
+\bra J_0;J_0 \ket_\textup{ret}(\omega)\right),
\end{equation} 
where 
$H_\textup{kin}=J\sum_{i=1}^{L}(S^x_iS^x_{i+1}+S^y_iS^y_{i+1})$ is the kinetic term
and $\bra \; ;\; \ket_\textup{ret}(\omega)$ is the retarded correlation function. 
The real part of the spin conductivity is written as
\begin{equation}
\Re\s(\omega)=\pi D(\beta) \delta(\omega) +\sigma_\textup{reg}(\beta, \omega),  
\end{equation}
where a finite Drude weight $D(\beta)>0$ implies an infinite dc conductivity
and $\sigma_\textup{reg}(\beta, \omega)$ represents the regular conductivity.  
Rewriting the spin conductivity $\sigma(\omega)$ by the current correlator and taking the zero frequency limit, we obtain
\begin{align}
D(\beta)&=\lim_{t \to \infty}\frac{\beta}{L}\bra J_0(t) J_0(0)\ket,
\nn\\
\mathcal{L}(\beta)
&:=\lim_{\omega \to 0}\s_{\textup{reg}}(\beta,\omega)
= 
\lim_{\tau\rightarrow\infty}\frac{\beta}{L}\int_{0}^\tau dt\left[\bra J_0(t) J_0(0)\ket-D(\beta)\right]. 
\label{L}\end{align} 
Here, $\bra\cdots \ket$ denotes the thermal average at inverse temperature $\beta$. 
In this form, $\mathcal{L}(\beta)$ is considered to be the next leading zero frequency contribution to the spin current correlation after $D(\beta)$. We call $\mathcal{L}(\beta)$ the spin dc conductivity. 

In the region of the anisotropy $0 \le \Delta <1$, non-zero $D(\beta)$ appears at finite temperatures \cite{Zotos}. 
The finite frequency contribution to the spin conductivity $\sigma_\textup{reg}(\beta, \omega)$ decays at the $\omega\to 0$ limit even at very high temperature in the periodic boundary conditions \cite{Brenes}. However, it is numerically suggested that the spin dc conductivity $\mathcal{L}(\beta)$ is a finite value for all anisotropy in the region $0 < \Delta <1$ \cite{Karrasch}.
This observation is also supported analytically by the Bethe ansatz method in a generic way \cite{ND diffusion},
that enables us to evaluate directly all the dc conductivities between any two conserved quantities
if the thermodynamic Bethe ansatz (TBA) equations for an integrable model are
grouped into the same fermionic type of the non-linear integral equations (NLIEs) as 
for the Lieb-Liniger model. 
Further, it allows us to relate the dc conductivities with the diffusion constants 
appearing in the Navier-Stokes equations by using the generalized hydrodynamic (GHD) theory (cf. \cite{Doyon 16, ND 16} for the Euler scale emergent hydrodynamics).        

In this paper, we concentrates on $\mathcal{L}(\beta)$, 
and we start from rewriting the general formula for the dc conductivities (see equation (4.19) in 
Ref.\cite{ND diffusion}) for the spin dc conductivity $\mathcal{L}(\beta)$ as follows:  
\begin{align}
\mathcal{L}(\beta)
&=
\frac{y_\alpha^2}{8\pi\beta}
\Bigg[\sum_{j=1}^{m_\alpha}\int d\la d\mu 
\frac{\eta_j(\la)}{(1+\eta_j(\la))^2}
\left\{\frac{\eta_{m_\alpha-1}(\mu)}{(1+\eta_{m_\alpha-1}(\mu))^2}+\frac{\eta_{m_\alpha}(\mu)}{(1+\eta_{m_\alpha}(\mu))^2}\right\}
\nn\\
&\qquad\times
\left|  
\left(\frac{\del_\la\ln\eta_j(\la)}{\del_{A\beta}\ln\eta_j(\la)}
-\frac{\del_\mu\ln\eta_{m_\alpha}(\mu)}{\del_{A\beta}\ln\eta_{m_\alpha}(\mu)}\right)
\frac{\partial_{A\beta} \ln\eta_j(\la)}{\partial_{A\beta} \ln\eta_{m_\alpha}(\mu)}\right| 
\left(T^\dr_{j,m_\alpha-1}(\la-\mu)\right)^2
\nn\\  
&\quad
+
\int d\la d\mu 
\left\{\frac{\eta_{m_\alpha-1}(\la)}{(1+\eta_{m_\alpha-1}(\la))^2}-\frac{\eta_{m_\alpha}(\la)}{(1+\eta_{m_\alpha}(\la))^2}\right\}
\nn\\
&\quad\hspace{3.8cm}\times
\left\{\frac{\eta_{m_\alpha}(\mu)}{(1+\eta_{m_\alpha}(\mu))^2}-\frac{\eta_{m_\alpha-1}(\mu)}{(1+\eta_{m_\alpha-1}(\mu))^2}\right\}
\nn\\
&\qquad\times
\left|  
\frac{\del_\la\ln\eta_{m_\alpha}(\la)}{\del_{A\beta}\ln\eta_{m_\alpha}(\la)}
-\frac{\del_\mu\ln\eta_{m_\alpha}(\mu)}{\del_{A\beta}\ln\eta_{m_\alpha}(\mu)}
\right| 
\left(T^\dr_{m_\alpha-1,m_\alpha-1}(\la-\mu)\right)^2\Bigg],
\label{LB}\end{align}
where $A=-2\pi J\sin\theta/\theta$.
The functions $\eta_j$ are the solutions to the TBA equations for the XXZ Hamiltonian (\ref{Hamiltonian}). 
In Section \ref{sec. 1/2 NLIEs},  we construct these equations based on the string assumption in which the lengths of the strings are restricted by the
Takahashi and Suzuki (TS) numbers $n_j \;(1 \le j,k \le m_\alpha)$ \cite{TS 72}. 
$T_{j,k}$ are the scattering kernels of strings and the superscript$\;^\dr$ represents the dressed quantities of the TBA \cite{Korepin}.
The number $y_\alpha$ represents the one particle magnetization of strings appearing in odd powers only on the final boundary strings as $\del_{2\beta h}\ln\eta_{m_\alpha}=\del_{2\beta h}\ln\eta_{m_\alpha-1}=y_\alpha/2$ and $\del_{2\beta h}\ln\eta_{j}=0 \;(1\le j<m_\alpha-1)$.

Our main result concerns the dressed scattering kernels. 
In Section \ref{sec. Tdr}, we obtain $T^\dr_{j,m_\alpha-1}$ in the high temperature limit 
by using the TBA equations for the integrable XXZ chain with arbitrary spin-$\s/2$ \cite{AS}. Let us  
write them down here:
\begin{align}
&T^\dr_{j, m_\alpha-1}(\la) 
\nn\\
&\; =\frac{\n_{j+1}}{y_r\n_{j+2}n_j}
\left\{
\sum_{s=1}^{n_{m_\alpha-1}-1}s\left(1-\frac{s}{n_{m_\alpha-1}}\right) \Delta a^{(m_\alpha-1)}_{j,s} (\la)
-
\sum_{s=1}^{n_{m_\alpha}-1}s\left(1-\frac{s}{n_{m_\alpha}}\right)\Delta a^{(m_{\alpha})}_{j,s} (\la)
\right\} 
\nn\\
&\qquad 
\for 
\quad m_r\le j<m_{r+1}, \quad j \ne m_\alpha-1,\; m_\alpha,    
\nn\\
&T^\dr_{m_\alpha-1, m_\alpha-1}(\la)
=
-T^\dr_{m_\alpha, m_\alpha-1}(\la)
\nn\\
&\; =
\frac{y_\alpha}{n_{m_\alpha-1}n_{m_\alpha}}
\left\{
\sum_{s=1}^{n_{m_\alpha-1}-1}s\left(1-\frac{s}{n_{m_\alpha-1}}\right)
+
\sum_{s=1}^{n_{m_\alpha}-1}s\left(1-\frac{s}{n_{m_\alpha}}\right)
\right\}
a(\la;2s),
\label{Intro Tdr}\end{align}
where
\begin{align}
\Delta a^{(j_\s)}_{j,s}(\la)
&:=
n_j a(\la; \tilde{q}_{j_\s}+\tilde{q}_{j+2}+2s)-\n_{j+2}a(\la; \tilde{q}_{j_\s}+q_j+2s),
\nn\\
a(\la;q)
&=
\frac{\theta}{2\pi}\frac{\sin\theta q}{\ch\theta \lambda-\cos\theta q}.
\label{Intro's a}\end{align}
The numbers $q_j$ and $\tilde{q}_j$ are the conjugate numbers of $n_j$ and the modified TS numbers $\n_j$ \cite{KSS, AS} respectively in the meaning that 
the sign $\pm$ changes and the center of the strings $\la$ shifts by $ip_0$ depending on the string parities $v_j$ (resp. $\tilde{v}_j$) in the one particle bare energies of strings as $a(\la;q_j)=-a\left(\la+i(1+v_j)p_0/2;n_j\right)$ (resp. $a(\la;\tilde{q}_j)=-a\left(\la+i(1+\tilde{v}_j)p_0/2; \n_j\right)$). The numbers $m_r$ and $y_r \; (1\le r \le \alpha)$ are uniquely determined 
from a given value of the anisotropy $\Delta$ and become the constituents of the TS numbers.
The string numbers, which are indicated by $j_\s$, relate with the number of the spin-$\s/2$ of the integrable chain whose TBA equations are used.   

In Section \ref{sec. high temp. L}, we calculate the high temperature limit of the spin dc conductivity $\mathcal{L}:=\lim_{\beta \to 0}\mathcal{L}(\beta)$ as a function of the anisotropy $\Delta$. 
We summarize our findings already here: $\mathcal{L}$ is discontinuous at rational numbers of $p_0=\pi/\cos^{-1}\Delta$---that is, if $y_\alpha$ increases in the way how $p_0$ approaches any rational number, $\mathcal{L}$ increases monotonically in proportion to $y_\alpha^2\ln y_\alpha$. The number $y_\alpha$ represents the magnetization as mentioned above.  
On the other hand, $\mathcal{L}$ is proportional to $y_\alpha$ when $\Delta$ approaches the isotropic point $(\Delta=1)$. This is surely higher than the lower bound on the spin diffusion constant---namely, the $\mathcal{L}$ divided by the thermal average of the magnetic susceptibility---which diverges logarithmically close to the isotropic point \cite{ND diffusion (0)}. 
We also found that $\mathcal{L}$ is proportional to $y_\alpha^2$ when $p_0$ approaches any irrational number. This agrees with the result obtained in the case 
where $p_0$ approaches the golden number \cite{Gopalakrishnan}.

\section{TBA equations for the spin-1/2 XXZ chain}\label{sec. 1/2 NLIEs}
Following the formulation of the TBA equations in \cite{TS 72} (see also \cite{T 99}),  
let us introduce the TS numbers $n_j$. These numbers represent the lengths of the strings which are formed by the spectral parameters $\la_k$ with $\la$ being their common real part as  
\begin{equation}
\la_k=\la+i(n_j+1-2k)+i\frac{(1-v_j)}{2}p_0, \quad k=1,2,\cdots n_j.
\end{equation}
The TS numbers $n_j$ are uniquely determined together with the parities $v_j \;(=\pm1)$ once given the anisotropy parameter $p_0$, and constituted by the series of numbers $m_r$ and $y_r$ as (\ref{TS no.}).  
When $p_0$ is a rational number in the region given by $0<1/p_0 \le1/2$, it can be 
expressed by a continued fraction with length $\alpha$ as follows: 
\begin{align}
&\frac{1}{p_0}=\frac{1|}{|\nu_1}+\frac{1|}{|\nu_2}+\cdots+\frac{1|}{|\nu_\alpha}
=\dfrac{1}{\nu_1+\dfrac{1}{\nu_2+\dfrac{1}{\dfrac{\ddots}{\nu_{\alpha-1}+\dfrac{1}{\nu_\alpha}}}}}\;, \nn\\
&\nu_2,\; \nu_3, \;\dots,\; \nu_{\alpha-1} \in \mathbb{N}, \quad \nu_1,\; \nu_{\alpha} \in \mathbb{N}_{\ge2}.
\label{continued fraction}\end{align}
For this rational number of $p_0$, $n_j$ satisfy the following closed relations:    
\begin{align}
&n_j=\frac{1}{2}\{(1-2\delta_{m_r,j})n_{j-1}+n_{j+1}\} \quad \for \quad m_r\le j \le m_{r+1}-2, \nn\\
&n_j=(1-2\delta_{m_{r-1},j})n_{j-1}+n_{j+1} \quad \for \quad j= m_r-1,\ r<\alpha, \nn\\
&n_0=0, \quad n_{m_\alpha-1}+n_{m_\alpha}=y_\alpha.
\label{n's relations}\end{align}
In the string excitations, the energy per site of the chain $e$ is given by  
\begin{align}
&e=\sum_j\int(\epsilon_j(\lambda)+2n_jh)\rho_j(\lambda)d\lambda-h, \nn\\
&\epsilon_j(\lambda)
=
-\frac{J\sin\theta\sin\theta n_j}{v_j\ch\theta \lambda-\cos\theta n_j}
=
-\frac{J\sin\theta\sin\theta q_j}{\ch\theta \lambda+\cos\theta q_j}
\qquad (1\le j \le m_\alpha),
\label{site energy}\end{align} 
where $\rho_j$ are the distribution functions of quasi-particles of strings.    
The one-particle dispersions $\epsilon_j$ are the derivatives with respect to the spectral parameter of the quasi-momenta of strings $\kappa_j$: 
\begin{align}
&\epsilon_j(\la)=\frac{J\sin\theta}{\theta}\frac{d}{d\la}\kappa_j(\la),    \nn\\
&\kappa_j(\la)=\pi-f(\la;n_j,v_j), \nn\\
&f(\la;n,v)=2v\tan^{-1}\left[\left\{\cot\left(\frac{n\theta}{2}\right)\right\}^v\tanh\left(\frac{\theta}{2}\la\right)\right] \nn\\
&\qquad\qquad=\pi-\frac{1}{i}\ln\left[\frac{\sh\left(\la+in+i\frac{(1-v)p_0}{2}\right)}{\sh\left(\la-in-i\frac{(1-v)p_0}{2}\right)}
\right].
\label{dispersion}\end{align}
In the expression (\ref{site energy}) for $\epsilon_j$, we used both $n_j$ and their
conjugate numbers $q_j$ that are determined by (\ref{TS no.}).
We rescale $\epsilon_j$ as   
\begin{align}
a_j(\la)&=A^{-1}\epsilon_j(\la)
\nn\\
&=
\frac{\theta}{2\pi}\frac{\sin\theta n_j}{v_j\ch\theta \lambda-\cos\theta n_j}
=
\frac{\theta}{2\pi}\frac{\sin\theta q_j}{\ch\theta \lambda+\cos\theta q_j}
\qquad 
(1\le j \le m_\alpha),
\end{align}
where
\begin{equation}
A:=-\frac{2\pi J\sin\theta}{\theta}. 
\label{rescaling}\end{equation}
These quantities are rewritten by (\ref{Intro's a}) as $a_j(\la)=-a\left(\la+ip_0;q_j\right)$ and satisfy the following relation in the form of a vector $\boldsymbol{a}=(a_j)$:
\begin{equation}
[\mathbbm{1}-\boldsymbol{S}*]\boldsymbol{a}(\la)=\mathbf{0},
\label{e's relations}\end{equation}
where $\mathbbm{1}$ is the identity matrix and the matrix $\boldsymbol{S}(\la)$ is defined 
by (\ref{matrix S}).  
We denote the convolution $\int d\mu\; b(\lambda-\mu)c(\mu)$ as $b*c(\la)$ for 
two arbitrary elements $b(\lambda)$ and $c(\lambda)$ in vectors and matrices.
From now on, we denote a series of arbitrary functions $\{f_j(\la)\}_{j=1}^{m_\alpha}$ by vector $\boldsymbol{f}=(f_j)$.   

The distribution functions $\rho_j$ and $\rho^\textup{h}_j$ of particles and holes of strings  satisfy the following integral relations:
\begin{align}
\varsigma_j(\rho_j(\la)+\rho_j^\textup{h}(\lambda))&=a_j(\lambda)- \sum_k\int d\mu\; T_{j,k}(\lambda-\mu)\rho_k(\mu) \nn\\
&=:a_j(\lambda)-\left[T*\rho\right]_j(\lambda),
\label{dressing ope.}\end{align}
where $\varsigma_j=\textup{sgn}(a_j)$.   
$T_{j,k}(\lambda)$ are the scattering kernels defined by:
\begin{align}
&T_{j,k}(\la)=\frac{1}{2\pi}\frac{d}{d\la} \Phi_{j,k}(\la), \nn\\
&\Phi_{j,k}(\la)=f(\la; |n_j-n_k|, v_jv_k)+f(\la; n_j+n_k, v_jv_k) \nn\\
                   &\hspace{4cm}  +2\sum_{i=1}^{\min(n_j,n_k)-1}f(\la; |n_j-n_k|+2i, v_jv_k). 
\end{align}  
These functions are symmetric as 
\begin{equation}
T_{j,k}(\lambda-\mu)=T_{j,k}(\mu-\lambda)=T_{k,j}(\lambda-\mu), 
\label{symmetry}\end{equation}
and satisfy the following relations in terms of 
$(T_{j,k})=(\boldsymbol{T}_1, \boldsymbol{T}_2, \cdots, \boldsymbol{T}_k, \cdots, \boldsymbol{T}_{m_\alpha})$:   
\begin{equation}
[\mathbbm{1}-\boldsymbol{S}*]\boldsymbol{T}_k(\la)=\boldsymbol{s}_k(\la),
\label{Tjk simple}\end{equation}
where $\boldsymbol{s}_k$ are the column vectors in the matrix $\boldsymbol{s}=(s_{j,k})=(\boldsymbol{s}_1, \cdots, \boldsymbol{s}_k, \cdots, \boldsymbol{s}_{m_\alpha})$ defined by (\ref{higher bare}).

At the state of the thermal equilibrium, we have the TBA equations determining \\
$\eta_j:=\rho_j^\textup{h}/\rho_j \; (1\le j \le m_\alpha-1)$ and $\eta_{m_\alpha}:=\rho_{m_\alpha}/\rho_{m_\alpha}^\textup{h}$:
\begin{align}
\ln\eta_j(\lambda)&=\beta g_j(\lambda)+\sum_k\int d\mu\; 
\varsigma_kT_{k,j}(\lambda-\mu)\ln(1+\eta_k^{-1}(\mu))
\nn\\
&=\beta g_j(\lambda) + \left[\varsigma T*\mathrm{ln}(1+\eta^{-1})\right]_j(\lambda), \nn\\
g_j(\lambda)&=Aa_j(\lambda)+2n_jh.
\label{TBA equations}\end{align}
Using (\ref{n's relations}), (\ref{e's relations}) and (\ref{Tjk simple}), these equations are rewritten as follows:  
\begin{equation}   
\boldsymbol{{\ln\eta}}(\la)=\beta\g(\la)+\boldsymbol{S}*\boldsymbol{\ln(1+\eta)}(\la), 
\label{TBA simple}\end{equation}
where the vectors $\g$ and $\boldsymbol{\ln(1+\eta)}$ are defined by (\ref{NLIEs for 1/2}).
Let us define the dressed energies $\varepsilon_j$, state densities $\rho_j^\textup{s}:=\rho_j+\rho_j^\textup{h}$ and Fermi weights $\vartheta_j$ as follows \cite{Korepin}:
\begin{align}
&\eta_j(\lambda)=e^{\beta\varepsilon_j(\lambda)}, \nn\\
&\vartheta_j(\lambda)=\frac{\rho_j(\lambda)}{\rho_j^\textup{s}(\lambda)} \quad (1 \le j \le m_\alpha-1),
\qquad \vartheta_{m_\alpha}(\lambda)=\frac{\rho_{m_\alpha}^\textup{h}(\lambda)}{\rho_{m_\alpha}^\textup{s}(\la)}.
\end{align}
Differentiating equations (\ref{TBA equations}) or (\ref{TBA simple}) with respect to $\beta$, we obtain $\varepsilon_j$ or $\boldsymbol{\varepsilon}$ as
\begin{align}
&\varepsilon_j(\lambda)=g(\lambda)-\left[\varsigma T*\vartheta\varepsilon\right]_j(\lambda),   \label{def. of dressing}\\
&\textup{or}\quad 
[\mathbbm{1}-\boldsymbol{S*(1-\vartheta)}]\boldsymbol{\varepsilon}(\la)=\g(\la), 
\label{dressed energy}\end{align}
where the matrix $\boldsymbol{S*(1-\vartheta)}$ is defined by (\ref{S with Fermi}).

\section{Dressed scattering kernels} \label{sec. Tdr}
As in the case of $\boldsymbol{\varepsilon}$,  
dressed quantities are obtained by differentiating some NLIEs, if available, with respect to the chemical potential, inverse temperature or any other parameter in general. 
Now suppose that there are NLIEs generating the following relations between the dressed scattering kernels $T^\dr_{j,k}$:   
\begin{align}
&T_{j,k}^\dr(\lambda)=T_{j,k}(\lambda)-\sum_l\int d\mu\; \varsigma_l T_{l,j}(\lambda-\mu)*\vartheta_l(\mu) T^\dr_{l,k}(\mu) \nn\\
&\hspace{1.1cm}=T_{j,k}(\lambda)-[\varsigma T*\vartheta T^\dr]_{j,k}(\lambda) 
\label{dressed Tjk}.
\end{align} 
Using (\ref{Tjk simple}), these relations are rewritten as: 
\begin{equation}
[\mathbbm{1}-\boldsymbol{S*(1-\vartheta)}]\boldsymbol{T}_k^\dr(\la)=\boldsymbol{s}_k(\la).   
\label{Tdr linear}\end{equation}
Comparing the dressed energy $\boldsymbol{\varepsilon}$ with $\boldsymbol{T}^{\dr}_k$, the former is just the energy of the one-particle excitation over a thermal state, obtained from the TBA equation (\ref{TBA simple}) via linear integral equation (\ref{dressed energy}) in which the driving term $\g$ exists. In this term, the first element $As_1$, defined by the function $s_r$ in (\ref{s, d}), plays the role of generating the kinetic energy and the final boundary element $y_\alpha h$ generates the magnetic energy respectively. 
On the other hand, 
the latter $\boldsymbol{T}^{\dr}_k$ are factors in the dc conductivities, the observable quantities in the diffusive dynamics.   
It was found in Ref.\cite{ND diffusion} that the two-particle excitation over a reference state contributes to the diffusive transports and three or higher particle excitations 
do not contribute to it. In the integrable models,
the reference should be taken from the generalized thermal state described by the ensemble constituted by an infinite set of conserved quantities---namely, the generalized Gibbs ensemble (GGE) \cite{Rigol}. 
The two-particle excitation is obtained by removing two particles with some momenta from the distribution of particles (i.e. making two holes with those momenta in the hole distribution) and by adding equal number of particles with different momenta. 
Note also that the quantities $\boldsymbol{T}^{\dr}_k$ are obtained via linear integral equations (\ref{Tdr linear}) whose driving terms $\boldsymbol{s}_k$ are composed of 
the rescaled kinetic energy elements $s_{r}$ defined by (\ref{higher bare}). From these facts, $\boldsymbol{T}^{\dr}_k$ have a clear physical interpretation, giving the rescaled energies of
the two-particle excitation.

In the case of the spin dc conductivity $\mathcal{L}(\beta)$, the two-particle excitation is constructed by the final boundary $n_{m_\alpha-1}$- and 
$n_{m_\alpha}$-strings, since only the dressed scattering kernels of these strings,  $T^\dr_{j,m_\alpha-1}=-T^\dr_{j,m_\alpha}$ do appear in the expression (\ref{LB}) for $\mathcal{L}(\beta)$---this relation between the final boundary kernels derives from the last line in (\ref{Tdr complicated}) and the symmetric property  (\ref{symmetry}) that is preserved after the dressing operation. To obtain $T^\dr_{j,m_\alpha-1}$, we first introduce the integrable XXZ chain with arbitrary spin-$\s/2$, and then we specifically choose the numbers of the spin as $\s+1=n_{m_\alpha-1}$ and $\s+1=n_{m_\alpha}=\n_{m_{\alpha-1}}$. 
In this course, 
let us define the particles and holes of strings with distribution functions $\rho^{(j_\s)}_j$ and $\rho^{\textup{h}(j_\s)}_j$ for the spin-$\s/2$ XXZ chain, which leads to 
the TBA equations determining  
$\eta^{(j_\s)}_j:=\rho_j^{\textup{h}(j_\s)}/\rho^{(j_\s)}_j \; (1\le j \le m_\alpha-1)$ and $\eta^{(j_\s)}_{m_\alpha}:=\rho^{(j_\s)}_{m_\alpha}/\rho_{m_\alpha}^{\textup{h}(j_\s)}$ \cite{AS}:
\begin{equation}
\boldsymbol{\ln\eta}\boldsymbol{\eta}^{(j_\s)}(\la)  
=\beta\g^{(j_\s)}(\la)+\boldsymbol{S}*\boldsymbol{\ln(1+\eta)}^{(j_\s)}(\la). 
\label{higher spin TBA simple}\end{equation}
Here, 
$j_\s$ indicates the string numbers satisfying the relation with the number of the spin as (\ref{spin number}) $\s+1=\tilde{n}_{j_\s}$.   
The vectors $\g^{(j_\s)}$ and $\boldsymbol{\ln(1+\eta)}^{(j_\s)}$ are defined by (\ref{NLIEs for s/2}).   
For the above specific numbers of the spin, the driving terms $\g^{(m_\alpha-1)}$ and $\g^{(m_{\alpha-1})}$ are related to $\boldsymbol{s}_{m_\alpha-1}$, defined already by (\ref{higher bare}), as follows: 
\begin{equation}
\del_A(\g^{(m_\alpha-1)}-\g^{(m_{\alpha-1})})
=\boldsymbol{s}_{m_\alpha-1}.
\label{generator}\end{equation}
We define the dressed energies $\varepsilon_j^{(j_\s)}$, state densities $\rho_j^{\textup{s}(j_\s)}$ and Fermi weights $\vartheta_j^{(j_\s)}$ as 
\begin{align}
&\eta_j^{(j_\s)}(\la)=e^{\beta\varepsilon_j^{(j_\s)}(\lambda)}, 
\qquad \rho^{\textup{s}(j_\s)}_j=\rho_j^{(j_\s)}+\rho^{\textup{h}(j_\s)}_j, \nn\\
&\vartheta_j^{(j_\s)}(\lambda)=\frac{\rho_j^{(j_\s)}(\lambda)}{\rho_j^{\textup{s}(j_\s)}(\lambda)} \quad (1 \le \j \le m_\alpha-1),
\qquad \vartheta_{m_\alpha}^{(j_\s)}(\lambda)=\frac{\rho_{m_\alpha}^{\textup{h}{(j_\s)}}(\lambda)}{\rho_{m_\alpha}^{\textup{s}{(j_\s)}}(\la)}.
\end{align}
Differentiating equation (\ref{higher spin TBA simple}) with respect to $\beta A$ and 
using (\ref{generator}), we obtain the following relation: 
\begin{align}
&\left[\mathbbm{1}-\boldsymbol{S*(1-\vartheta)}^{(m_\alpha-1)}\right]\del_A\boldsymbol{\varepsilon}^{(m_\alpha-1)}(\la) \nn\\
&\qquad -
\left[\mathbbm{1}-\boldsymbol{S*(1-\vartheta)}^{(m_{\alpha-1})}\right]\del_A\boldsymbol{\varepsilon}^{(m_{\alpha-1})}(\la) 
=\boldsymbol{s}_{m_\alpha-1}(\la),
\label{higher relations}\end{align} 
where the matrix $\boldsymbol{S*(1-\vartheta)}^{(j_\s)}$ is defined by (\ref{higher S with Fermi}).

\subsection*{The $T\rightarrow \infty$ limit of $T^{\mathrm{dr}}_{j,m_\alpha-1}$}
In the high temperature limit, all the Fermi weights $\vartheta^{(j_\s)}_j(\la)=1/(1+\eta^{(j_\s)}_j(\la))$ are constant with respect to the spectral parameter $\la$ since the driving term $\beta\g^{(j_\s)}(\la)$ vanishes from equation (\ref{higher spin TBA simple}),
leaving only constant and common $\boldsymbol{\eta}^{(j_\s)}$ for all $j_\s $---that is,
$
\boldsymbol{\eta}^{(j_\s)}=\boldsymbol{\eta}^{(2)}=\boldsymbol{\eta}$ for 
$2\le j_\s \le m_\alpha$.   Therefore, relation (\ref{higher relations}) reduces to  
\begin{align}
[\mathbbm{1}-\boldsymbol{S*(1-\vartheta)}]
\del_A\left\{\boldsymbol{\varepsilon}^{(m_\alpha-1)}(\la)-\boldsymbol{\varepsilon}^{(m_{\alpha-1})}(\la)\right\} =\boldsymbol{s}_{m_\alpha-1}(\la)+\boldsymbol{O}(\beta).
\end{align}
Comparing this relation with (\ref{Tdr linear}), we obtain the dressed scattering kernels 
$\boldsymbol{T}^\dr_{m_\alpha-1}$ as
\begin{align}
&\boldsymbol{T}^\dr_{m_\alpha-1}(\la)
=\del_A\left\{\boldsymbol{\varepsilon}^{(m_\alpha-1)}(\la)
-\boldsymbol{\varepsilon}^{(m_{\alpha-1})}(\la)\right\} +\boldsymbol{O}(\beta).
\label{formula of Tdr}\end{align}
Using the high temperature expansions of $\eta_j^{(j_\s)}$ in (\ref{limit of eta}),
we obtain the rescaled energies $\del_A\varepsilon^{(j_\s)}_j$ for all $j_\s$ as follows:
\begin{align}
\del_A\varepsilon_j^{(j_\s)}(\la)
&=
\del_{\beta A}\ln\eta^{(j_\s)}_j(\la)
\nn\\ &=
\frac{\n_{j+1}}{y_r\n_{j+2}n_j}
\sum_{s=1}^{\s}\frac{s(\s+1-s)}{\s+1} \Delta a^{(j_\s)}_{j,s} (\la) 
\nn\\
\for \quad& m_r\le j < m_{r+1}, \quad j \ne m_\alpha-1, \; m_\alpha, 
\nn\\
\del_A\varepsilon_{m_\alpha-1}^{(j_\s)}(\la)
&=
-\del_A\varepsilon_{m_\alpha}^{(j_\s)}(\la)=\del_{\beta A}\ln\eta^{(j_\s)}_{m_\alpha-1}(\la) 
\nn\\ &=
\frac{y_\alpha}{n_{m_\alpha-1}n_{m_\alpha}}
\sum_{s=1}^{\s}\frac{s(\s+1-s)}{\s+1} 
a_{m_\alpha-1,s}^{(j_\s)}(\la),
\label{del epsilon}\end{align}
where  
$
\Delta a^{(j_\s)}_{j,s}(\la)
=
n_j a(\la; \tilde{q}_{j_\s}+\tilde{q}_{j+2}+2s)-\n_{j+2}a(\la; \tilde{q}_{j_\s}+q_j+2s)
$ 
as defined by the function 
$a(\la;q)=\frac{\theta}{2\pi}\frac{\sin\theta q}{\ch\theta \lambda-\cos\theta q}$
in (\ref{Intro's a}), and
\begin{align}
a_{j,s}^{(j_\s)}(\la)
&:=
\frac{\theta}{2\pi}\frac{\sin\theta q_{j,s}^{(j_\s)}}{\ch\theta \lambda+\cos\theta q_{j,s}^{(j_\s)}},
\nn\\
q_{j,s}^{(j_\s)}
&\equiv 
\tilde{q}_{j_\s}+q_j+2s-\frac{1+\tilde{v}_{j_s}v_j}{2}p_0 \mod 2p_0.
\label{higher a}\end{align}
The parities $v_j, \; \tilde{v}_j$ and the conjugate numbers $q_j,\; \tilde{q}_j$ of the TS numbers and their modified ones $n_j,\; \n_j$ are determined 
by (\ref{TS no.}) and (\ref{modified TS no.}) respectively.  
From (\ref{formula of Tdr}) and (\ref{del epsilon}), we obtain the $T \to \infty$ limit values of $T^\dr_{j,m_\alpha-1}$ as (\ref{Intro Tdr}), which we again show here and rewrite as follows by using the relation $q_{m_\alpha-1}=-q_{m_\alpha}$ to reduce the sums over the final boundary TS numbers:  
\begin{align}
&T^\dr_{j, m_\alpha-1}(\la) 
\nn\\
&\;=
\frac{\n_{j+1}}{y_r\n_{j+2}n_j}
\left\{
\sum_{s=1}^{n_{m_\alpha-1}-1}s\left(1-\frac{s}{n_{m_\alpha-1}}\right) \Delta a^{(m_\alpha-1)}_{j,s} (\la)
-
\sum_{s=1}^{n_{m_\alpha}-1}s\left(1-\frac{s}{n_{m_\alpha}}\right)\Delta a^{(m_{\alpha})}_{j,s} (\la)
\right\} 
\nn\\
&\;=
\frac{\n_{j+1}y_\alpha}{\n_{j+2}n_{m_\alpha-1}n_{m_\alpha}}K_j^\dr(\la)
\qquad  \for 
\quad m_r\le j<m_{r+1}, \quad j \ne m_\alpha-1,\; m_\alpha,    
\nn\\
&T^\dr_{m_\alpha-1, m_\alpha-1}(\la)=-T^\dr_{m_\alpha, m_\alpha-1}(\la) 
\nn\\
&\;=
\frac{y_\alpha}{n_{m_\alpha-1}n_{m_\alpha}}
\left\{
\sum_{s=1}^{n_{m_\alpha-1}-1}s\left(1-\frac{s}{n_{m_\alpha-1}}\right)
+
\sum_{s=1}^{n_{m_\alpha}-1}s\left(1-\frac{s}{n_{m_\alpha}}\right)
\right\}
a(\la;2s)
\nn\\
&\;=
\frac{y_\alpha}{n_{m_\alpha-1}n_{m_\alpha}}K_{m_\alpha-1}^\dr(\la),
\label{T}\end{align}
where 
\begin{align}
K^\dr_j(\la) 
&:=
\sum_{s=1}^{n_j-1}
\left(\frac{2s(n_j-s)}{n_j}+y_r\right)
a(\la; q_{m_\alpha}+q_j+2s)
\nn\\
&\quad
+\sum_{s=1}^{y_r}\frac{s^2}{y_r}
\bigg\{
a(\la; q_{m_\alpha}+\tilde{q}_{j+2}+2s)
+a(\la; q_{m_\alpha}-\tilde{q}_{j+2}-2s)
\bigg\} 
\nn\\
\for & 
\quad m_r\le j<m_{r+1}  \;(r \le \alpha-2)  
\;\; \textup{and} \;\; 
j=m_{\alpha-1} \;(r=\alpha-1),  
\nn\\
K^\dr_j(\la) 
&:=
\frac{2n_{m_\alpha}}{n_j}\sum_{s=1}^{\left[\frac{n_{j-1}-1}{2}\right]}
(n_{j-1}-2s)a(\la; q_{j-1}+2s)
\nn\\
&\quad
+\sum_{s=1}^{n_{m_\alpha}}
\Bigg\{\left(2s+n_{m_\alpha}-\frac{\n_{j+2}s^2}{n_jn_{m_\alpha}} \right)
a(\la; q_{j+1}+2s)
+
\frac{s^2}{n_{m_\alpha}}  
a(\la; q_{m_\alpha}+\tilde{q}_{j+2}+2s)
\Bigg\}
\nn\\
\for & 
\quad m_{\alpha-1} < j \le m_\alpha-2,  
\nn\\
K^\dr_{m_\alpha-1}(\la)&:=-K^\dr_{m_\alpha}(\la) 
\nn\\
&=
n_{m_\alpha}\sum_{s=0}^{\left[\frac{y_\alpha}{2}\right]-n_{m_\alpha}}
\left(\frac{n_{m_\alpha}+2s}{n_{m_\alpha-1}}-1\right)a(\la;2q_{m_\alpha}-2s)
\nn\\
&\quad
+\sum_{s=1}^{n_{m_\alpha}-1}s\left(2-\frac{y_\alpha s}{n_{m_\alpha-1}n_{m_\alpha}}\right)a(\la;2s).
\label{K}\end{align}
Here, $[x]$ denotes the maximum integer less than or equal to $x$ (Gauss's symbol).

\section{The $T \rightarrow \infty $ limit of $\mathcal{L}(\beta)$} \label{sec. high temp. L}
\begin{figure}[H]
    \centering
    \includegraphics[width=1\columnwidth]{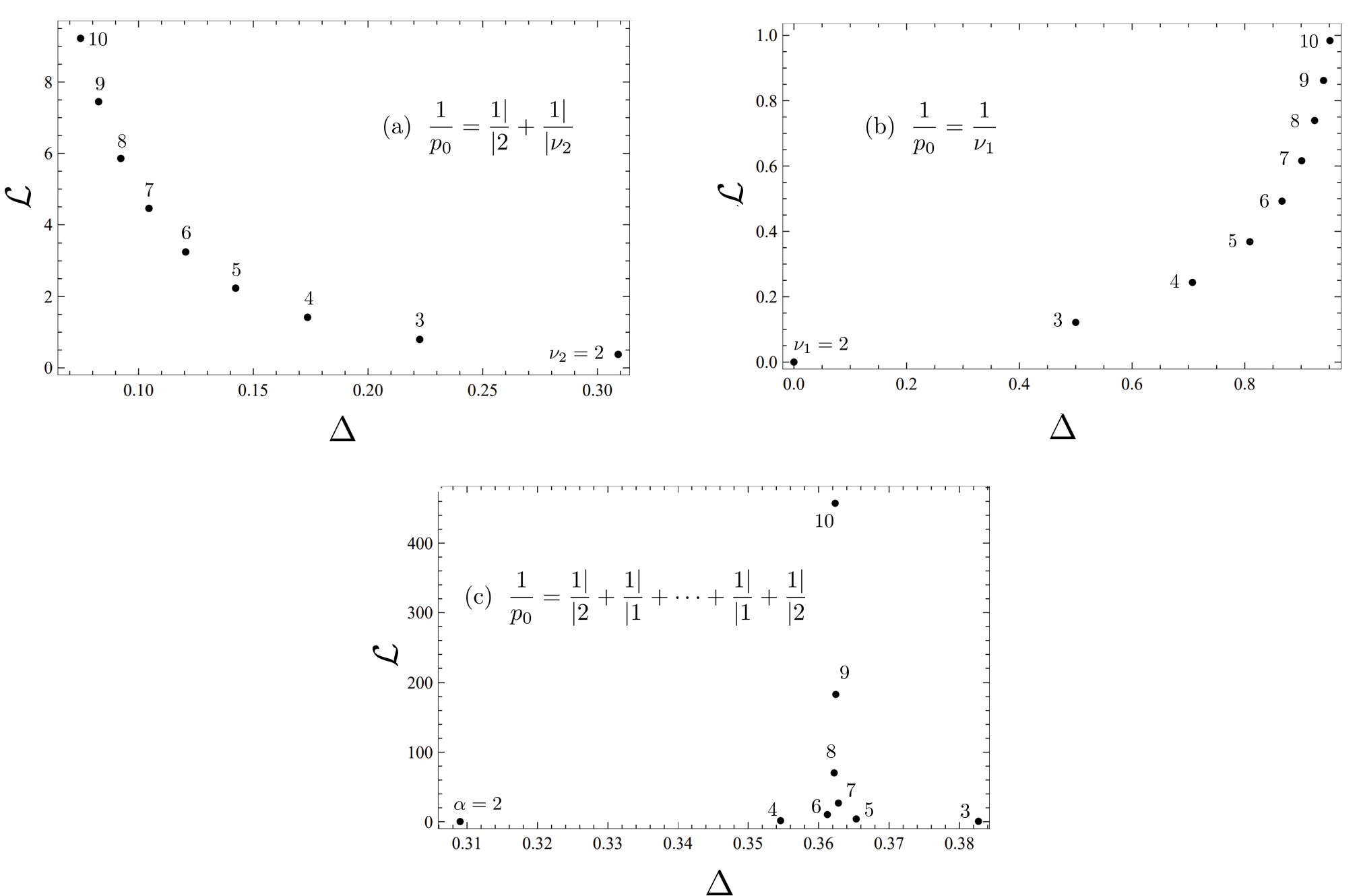}
\caption{$\mathcal{L}$ evaluated for various anisotropy parameters;
(a) $\frac{1}{p_0}=\frac{1|}{|2}+\frac{1|}{|\nu_2} \;\; (\nu_2=2,\cdots,10)$,\\
(b) $\frac{1}{p_0}=\frac{1}{\nu_1}\;\; (\nu_1=2,\cdots, 10)$
and (c) $\frac{1}{p_0}=\frac{1|}{|2}+\frac{1|}{|1}+\cdots +\frac{1|}{|1}+ \frac{1|}{|2} \;\; (2 \le \alpha \le 10)$.\\
The points ($\bullet$) indicate exact values from (\ref{spin dc}) and (\ref{L terms}) with coupling constant $J=1$. 
}  
\label{Fig. L's}  
\end{figure}

\begin{figure}[H]
    \centering
    \includegraphics[width=0.5\columnwidth]{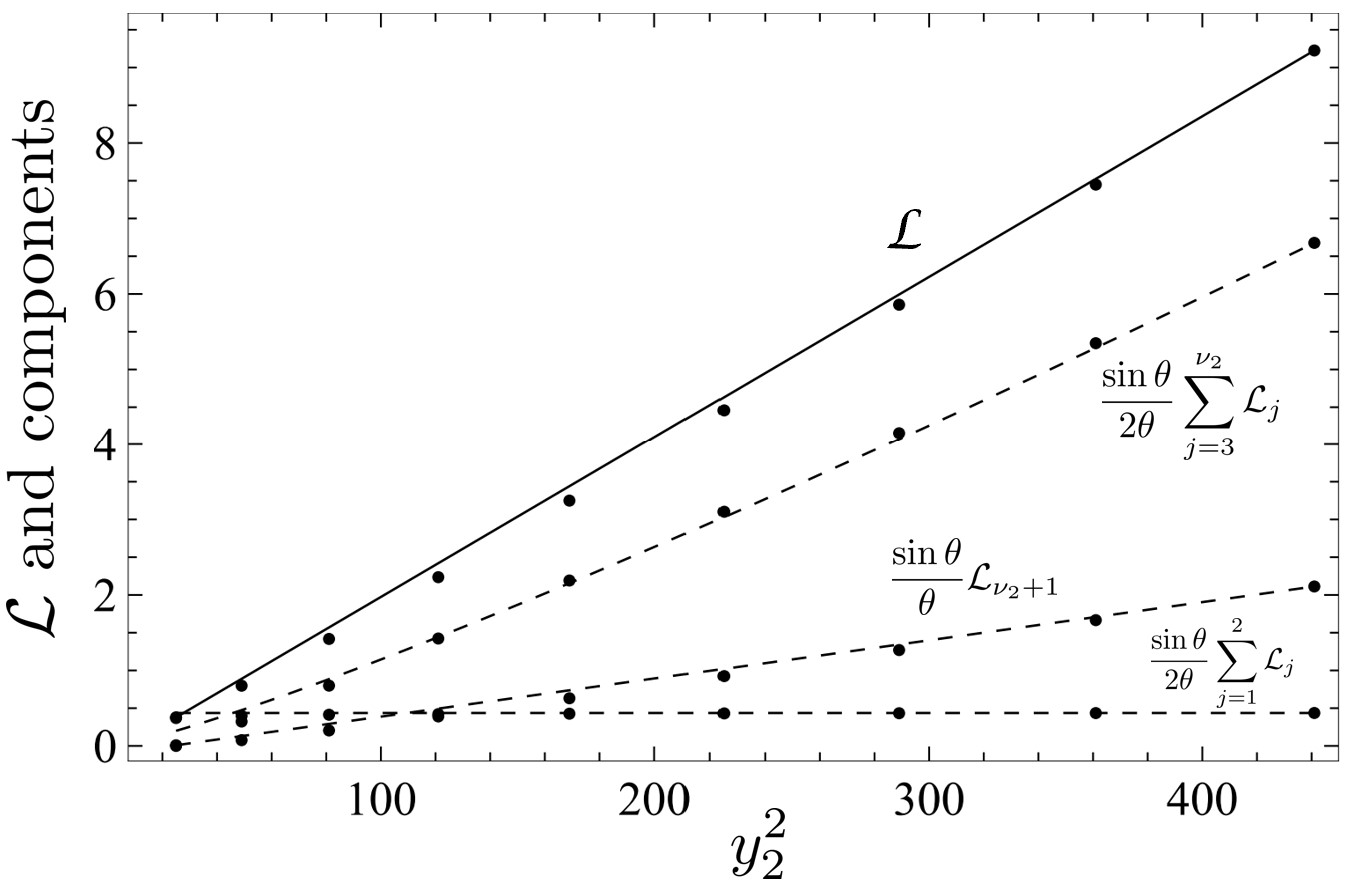}
\caption{$\mathcal{L}$ for $\frac{1}{p_0}=\frac{1|}{|2}+\frac{1|}{|\nu_2} \; (\nu_2=2,\cdots,10)$ as a function of $y_2=1+2\nu_2$ . 
The continuous and dashed lines correspond to the total and partial spin dc conductivities
respectively. The straight lines connect the endpoints at $\nu_2=2$ and 10. The curved line for the component $\sum_{j=3}^{\nu_2}\mathcal{L}_j$ is given by $(5.0\times10^{-3})y_2^2\ln y_2$.
}  
\label{alpha=2}  
\end{figure}

\begin{figure}[H]
    \centering
    \includegraphics[width=0.5\columnwidth]{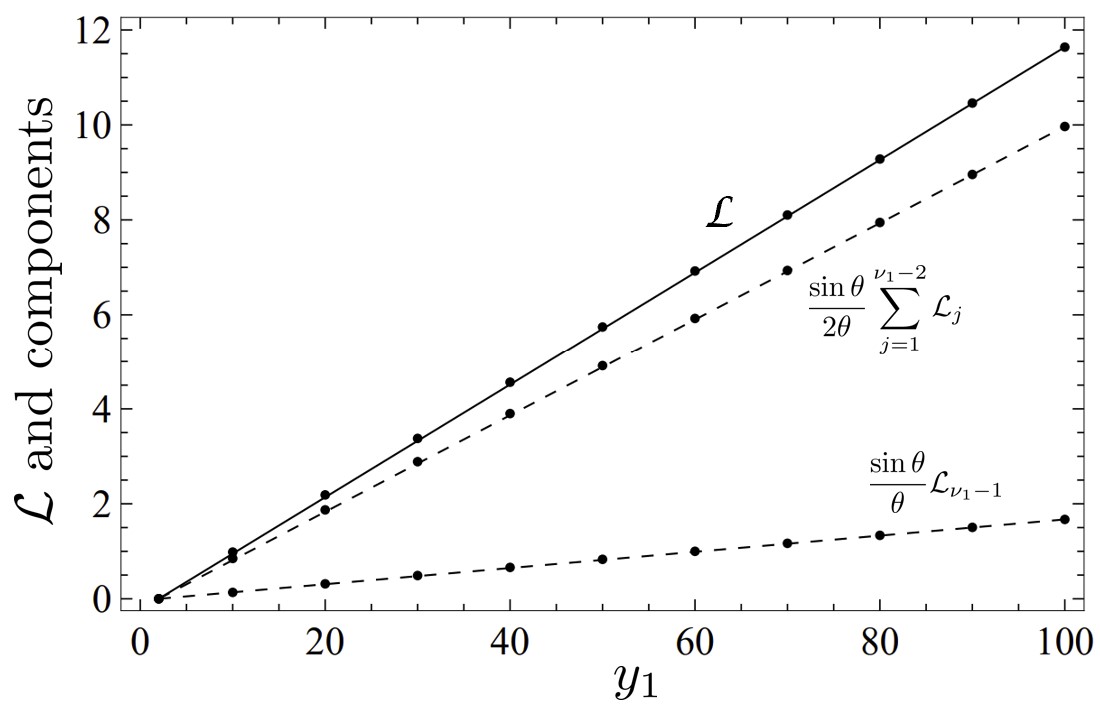}
\caption{$\mathcal{L}$ for $\frac{1}{p_0}=\frac{1}{\nu_1} \; (\nu_1=2, 10, 20, \cdots,100)$ as a function of $y_1=\nu_1$. The lines connect the points at $\nu_1=2$ and 100. 
 } 
\label{alpha=1}  
\end{figure}

\begin{figure}[H]
    \centering
    \includegraphics[width=0.5\columnwidth]{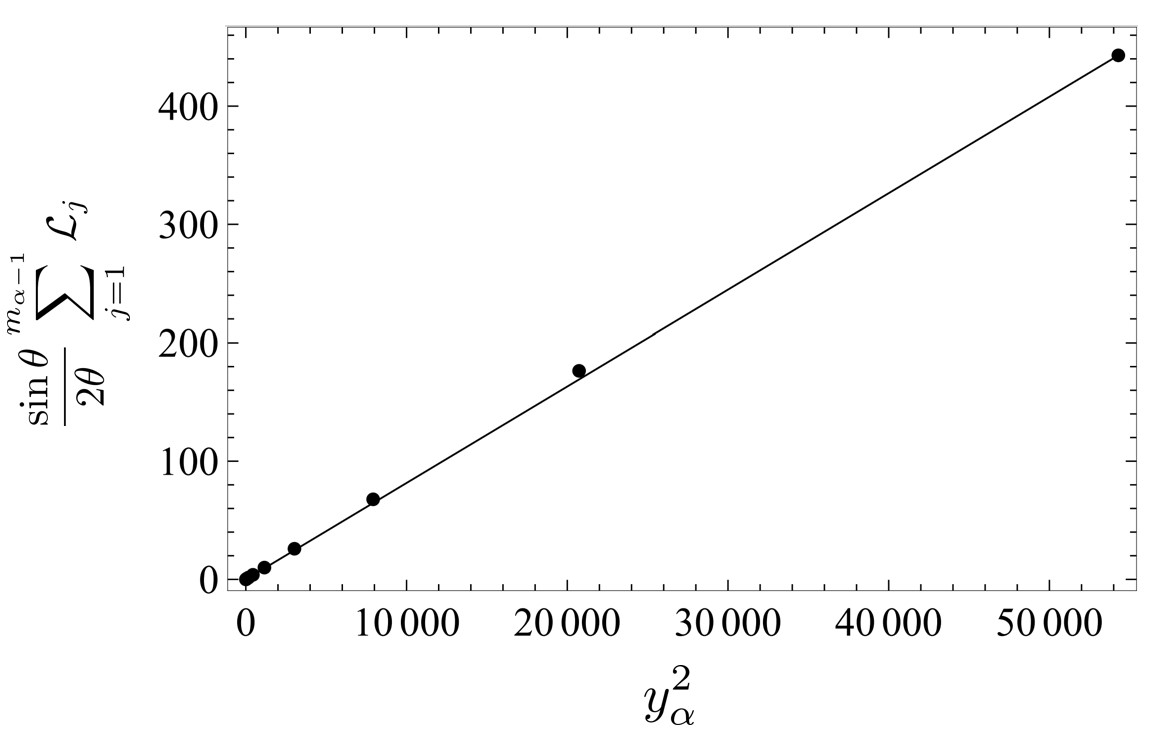}
\caption{
$(\sin\theta/2\theta)\sum_{j=1}^{m_{\alpha-1}}\mathcal{L}_j$ for
$\frac{1}{p_0}=\frac{1|}{|2}+\frac{1|}{|1}+\cdots +\frac{1|}{|1}+ \frac{1|}{|2} \;(1 \le \alpha \le 10)$ as a function of $y_\alpha$. \\
The lines connect the points at $\alpha=1$ and 10.
} 
\label{αto infty}  
\end{figure}

In the high temperature limit, the relation between the final boundary solutions to the NLIEs (\ref{TBA simple}) reduces to $\eta_{m_\alpha-1}\eta_{m_\alpha}=1$. Thus, the spin dc conductivity (\ref{LB}) reduces to the following limit value $\mathcal{L}$:  
\begin{align}
\mathcal{L}&:=\lim_{\beta\to0}\mathcal{L}(\beta)
\nn\\
&=
\frac{y_\alpha^2}{4\pi\beta}
\sum_{j=1}^{m_\alpha}\int d\la d\mu 
\frac{\eta_j(\la)}{(1+\eta_j(\la))^2}
\frac{\eta_{m_\alpha}(\mu)}{(1+\eta_{m_\alpha}(\mu))^2}
\nn\\
&\quad \times
\left|
\left(
\frac{\del_\la\ln\eta_j(\la)}{\del_{A\beta}\ln\eta_j(\la)}
-\frac{\del_\mu\ln\eta_{m_\alpha}(\mu)}{\del_{A\beta}\ln\eta_{m_\alpha}(\mu)}
\right)
\frac{\partial_{A\beta} \ln\eta_j(\la)}{\partial_{A\beta} \ln\eta_{m_\alpha}(\mu)}
\right| 
\left(T^\dr_{j,m_\alpha-1}(\la-\mu)\right)^2.  
\label{L at inifinite temp}\end{align}
Using the high temperature expansions of $\eta_j=\eta_j^{(2)}$ in (\ref{limit of eta}) 
and substituting the expressions (\ref{T}) and (\ref{K}) for the dressed scattering kernels, we obtain 
\begin{equation}
\mathcal{L}
=
\frac{\left|A\right|}{4\pi}
                   \left(\sum_{j=1}^{m_\alpha-2}\mathcal{L}_j+2\mathcal{L}_{m_\alpha-1}\right),  
\label{spin dc}\end{equation}
where
\begin{align}
\mathcal{L}_j
&=\frac{y_ry_\alpha}{\n_{j+1}\n_{j+2}^2} 
\int d\la d\mu
\left|(\del_\la+\del_\mu)\frac{\Delta a_{j,1}^{(2)}(\la)}{a_{m_\alpha}(\mu)}\right|    
(K^\dr_j(\la-\mu))^2 \nn\\
&\for 
\quad 
m_r\le j<m_{r+1}, \quad j\ne m_{\alpha-1}, \; m_\alpha,  
\nn\\
\mathcal{L}_{m_\alpha-1}
&= 
\int d\la d\mu 
\left|(\del_\la+\del_\mu)\frac{a_{m_\alpha}(\la)}{a_{m_\alpha}(\mu)}\right| 
(K^\dr_{m_\alpha-1}(\la-\mu))^2.  
\label{L terms}\end{align}
In Fig.\ref{Fig. L's}-(a), (b) and (c), $\mathcal{L}$ is shown as a function of $\Delta$ for 
different characteristic anisotropy parameters expressed by $1/p_0$. We set the coupling constant $J=1$.   
When the last element $\nu_\alpha$ of the continued fraction (\ref{continued fraction}) increases, $p_0$ approaches a rational number.  
In the case of (a) and (b), the anisotropy approaches the free fermion point ($\Delta=0$)
and isotropic point ($\Delta=1$) respectively. 
We find that $\mathcal{L}$ is a monotonically increasing function of $\nu_\alpha$.
On the other hand when the length of the continued fraction $\alpha$ increases, $p_0$ approaches an irrational number.
In the case of (c), $p_0$ approaches $1+(1+\sqrt{5})/2$, where $(1+\sqrt{5})/2$ is the golden number. 
We find that $\mathcal{L}$ is a monotonically increasing function of $\alpha$.  
To check these behaviors,   
We show $\mathcal{L}$ and its components (\ref{L terms}) as a function of $y_\alpha$.
By definition (\ref{p, m, y}), this number increases with $\nu_\alpha$ and $\alpha$,  
and this number represents the one particle magnetization of strings as mentioned in Section \ref{Intro. and Summ.}. 
In Fig.\ref{alpha=2}, the anisotropy approaches the $\Delta=0$ point. 
At this point, the spin dc conductivity vanishes ($\mathcal{L}=0$) as shown in Fig.\ref{Fig. L's}-(b). However, $\mathcal{L}$ increases by more than the second power of $y_2$.  
In Fig.\ref{alpha=1}, it approaches the $\Delta=1-$ point. In this case $\mathcal{L}$ is proportional to $y_1 \sim (1-\Delta)^{-1/2}$.
This coincides with the result obtained in the case of the gapped spin-1/2 XXZ chain
with the anisotropy close to the $\Delta=1+$ point. In that case, the spin diffusion constant---namely, the $\mathcal{L}$ divided by the magnetic susceptibility at the equilibrium--- is proportional to $(\Delta-1)^{-1/2}$ in the high temperature limit \cite{ND diffusion, Gopalakrishnan2}. 
In Fig.\ref{αto infty}, $p_0$ approaches $1+\frac{1+\sqrt{5}}{2}$.   
In this case $(\sin\theta/2\theta)\sum_{j=1}^{m_{\alpha-1}}\mathcal{L}_j$ is proportional to the second power of $y_\alpha$. 
Since irrational numbers are given by endless continued fractions, the last term $(\sin\theta/\theta)\mathcal{L}_{m_\alpha-1}$ was excluded.    

We examine the value $\mathcal{L}$ to obtain the leading order terms in the power series of $y_\alpha$.
First, the derivatives in the expressions (\ref{L terms}) for $\mathcal{L}_j$ are scaled as 
\begin{align}
\left|(\del_\la+\del_\mu)\frac{\Delta a_{j,1}^{(2)}(\la)}{a_{m_\alpha-1}(\mu)}\right|    
&=
\frac{\theta}{\sin\frac{\pi}{y_\alpha}}
\Bigg|
\n_{j+2}\sin\theta q_j\frac{\sh\theta(\la-\mu)+\cos\frac{\pi}{y_\alpha}\sh\theta\la-\cos\theta q_j\sh\theta \mu}{(\ch\theta\la+\cos\theta q_j)^2} 
\nn\\
&\qquad\qquad\quad-
n_j\sin\theta \tilde{q}_{j+2}
\frac{\sh\theta(\la-\mu)+\cos\frac{\pi}{y_\alpha}\sh\theta\la-\cos\theta \tilde{q}_{j+2}\sh\theta \mu}{(\ch\theta\la+\cos\theta \tilde{q}_{j+2})^2}
\Bigg|
\nn\\
&\sim
\frac{\theta}{\pi/y_\alpha}
|\tilde{n}_{j+2}\sin\theta q_j-n_j\sin\theta \tilde{q}_{j+2}|
\nn\\
&\sim
\theta \nu_{r+1}x_{\alpha-1-r}y_r
\qquad  
\for \quad m_r\le j< m_{r+1}, \quad j \ne m_{\alpha}-1,\; m_{\alpha}, 
\nn\\
\left|(\del_\la+\del_\mu)\frac{a_{m_\alpha-1}(\la)}{a_{m_\alpha-1}(\mu)}\right|
&\sim
\theta
\qquad 
\for \quad j=m_\alpha-1,
\label{scale}\end{align}  
where $a \sim b$ means that $\lim_{y_\alpha\to\infty}a/b \sim 1$ for two arbitrary values $a$ and $b$.
To obtain the third line in (\ref{scale}), we used that 
$n_j=y_{r-1}+(j-m_r)y_r$ for $m_r \le j < m_{r+1}$ 
from (\ref{TS no.}) and the following equality from (\ref{new q}) and (\ref{p's proof}):
\begin{equation} 
|\theta q_j| =\{x_{(\alpha-r-1)-1}+(m_\alpha-j-l_{\alpha-r-1})x_{\alpha-r-1}\}\frac{\pi}{y_\alpha}
\qquad \for \quad  m_r \le j < m_{r+1}.
\end{equation}
The modified numbers, $\n_j$ and $|\theta\tilde{q}_j|$ are treated in the same way from (\ref{modified TS no.}). 
Second, in order to scale the $K_j^\dr$ expressed by (\ref{K}), 
we consider the expansion of the scaled bare energy in a power series of $|\theta q_{m_\alpha}|=\pi/y_\alpha$:
\begin{equation}
a(\la; q_{m_\alpha}+q)
=
a^{(0)}(\la;q)+\frac{\pi}{y_\alpha}a^{(1)}(\la;q)+\left(\frac{\pi}{y_\alpha}\right)^2a^{(2)}(\la;q)+\cdots.
\label{a's expansion}\end{equation}
All the zeroth order terms $a^{(0)}(\la;q)=a(\la;q)$ disappear from the first component $\{K_j^\dr\}_{j=1}^{m_{\alpha-1}}$, where $q=q_j+2s$ or $q=\pm(\tilde{q}_{j+2}+2s)$. This is because these terms are canceled out of the sums over the TS numbers
by rewriting the terms as $a(\la;q_j+2s)=-a(\la+i(1+v_j)p_0/2;n_j-2s)$ and using the relation 
$a(\la;-n)=-a(\la;n)$.
On the other hand the zeroth order terms do not disappear from the remaining 
component $\{K_j^\dr\}_{j=m_{\alpha-1}+1}^{m_\alpha-1}$.
Thus, $K_j^\dr$ are scaled as  
\begin{align}
K^\dr_j(\la)
&\sim
\frac{\pi}{y_\alpha}
\left\{
\sum_{s=1}^{n_j-1}\left(\frac{2s(n_j-s)}{n_j}+y_r\right)
+\sum_{s=1}^{y_r}\frac{2s^2}{y_r}
\right\}a^{(1)}(\la;q_j+2s)
\nn\\
&\for \quad m_r\le j<m_{r+1}  \;(r \le \alpha-2) \;\; \textup{and} \;\;
j=m_{\alpha-1} \;(r=\alpha-1),
\nn\\
K^\dr_j(\la)
&\sim
\left\{   
\frac{2n_{m_\alpha}}{n_j}\sum_{s=1}^{\left[\frac{n_{j-1}-1}{2}\right]}
(n_{j-1}-2s)
+
\sum_{s=1}^{n_{m_\alpha}}
\left(2s+n_{m_\alpha}-\frac{2s^2}{n_j}\right)
\right\}a(\la; q_{j-1}+2s)
\nn\\
&\for \quad m_{\alpha-1}< j  \le m_{\alpha}-2,
\nn\\
K^\dr_{m_\alpha-1}(\la)
&\sim
\left\{
n_{m_\alpha}\sum_{s=0}^{\left[\frac{y_\alpha}{2}\right]-n_{m_\alpha}}
\left(1-\frac{n_{m_\alpha}+2s}{n_{m_\alpha-1}}\right)
+
\sum_{s=1}^{n_{m_\alpha}-1}s\left(2-\frac{y_\alpha s}{n_{m_\alpha-1}n_{m_\alpha}}\right)\right\}a(\la;2s).
\label{scale 2}\end{align}

\vspace{3mm}\noindent$\bullet$\; 
\underline{For $\nu_r \sim 1 \; (1 \le r<\alpha)$ and $1\ll \nu_\alpha < \infty$}\;; 
in conjunction with (\ref{L terms}), (\ref{scale}) and (\ref{scale 2}),
we find that all the $\mathcal{L}_j$ for $1\le j \le m_{\alpha-1}$ are scaled  as $\mathcal{L}_j \sim 1$.
This is obtained from $\theta \sim 1$, $y_r \sim n_j \sim 1$ and 
$\nu_{r+1}x_{\alpha-1-r}\sim  y_\alpha$, in which the last ordering is used in (\ref{scale}).  
To scale the second component $\{\mathcal{L}_j\}_{j=m_{\alpha-1}+1}^{m_{\alpha}-2}$, we use that 
$
a(\la;q_j+2s)
=
a\left(\la+i\frac{1-v_jv_{j-s^\prime}}{2}p_0; q_{j-s^\prime}+s^{\prime\prime}\right)
$,
where
$2s=n_{m_\alpha}s^\prime + s^{\prime\prime}$. 
From this, we obtain   
\begin{align}
K^\dr_j 
&\sim
\frac{1}{n_j}\sum_{s=1}^{j-1-m_{\alpha-1}}\sum_{s^\prime=1}^{\left[\frac{n_{m_\alpha}}{2}\right]}(n_{j-s}-2s^\prime)
a\left(\la+i\frac{1-v_{j-1}v_{j-s}}{2}p_0; q_{j-s}+2s^\prime\right)
\nn\\
&\sim
\frac{1}{n_j}\sum_{s=1}^{j-1-m_{\alpha-1}}n_{j-s}
a(\la; q_{j-s})
\sim
\frac{1}{n_j}\sum_{s=1}^{j-1-m_{\alpha-1}}n_{s+m_{\alpha-1}}
a(\la; q_{s+m_{\alpha-1}})
\nn\\
\for & \quad m_{\alpha-1}< j  \le m_{\alpha}-2,
\end{align}
where we assumed that $\nu_\alpha$ is sufficiently large to allow us 
to neglect the number $n_{m_\alpha}$. Thus we obtain
$
\sum_{j=m_{\alpha-1}+1}^{m_{\alpha}-2}\mathcal{L}_j
\sim 
\sum_{j=1}^{\nu_\alpha-2}\frac{y_\alpha\nu_\alpha}{(j+1)(j+2)^2}
\left[
\frac{1}{j}\sum_{s=1}^{j-1}s \frac{\nu_\alpha-s}{y_\alpha}\right]^2
\sim
y_\alpha^2\ln y_\alpha. 
$
The final boundary term is scaled easily as $\mathcal{L}_{m_\alpha-1} \sim y_\alpha^2$. In total, the spin dc conductivity in the high temperature limit is scaled as follows when $p_0$ approaches rational numbers by large $\nu_\alpha \; (\alpha >1)$:
\begin{equation}
\mathcal{L} \sim  y_\alpha^2 \ln y_\alpha.
\label{total scale}\end{equation} 

\vspace{3mm}\noindent$\bullet$\; 
\underline{For $1\ll \nu_1 < \infty$}\;; 
using that  
$
a(\la;q_j+2s)
=
-a(\la+ip_0; j-1-2s)$, 
we obtain 
$\sum_{j=1}^{y_1-2}\mathcal{L}_j\sim y_1$. 
Since the final boundary term is also scaled as  
$\mathcal{L}_{y_1-1} \sim y_1$, 
$\mathcal{L}$ is scaled as follows when $p_0$ approaches the isotropic $\Delta=1$ point:
\begin{equation}
\mathcal{L}\sim y_1. 
\end{equation}

\vspace{3mm}\noindent$\bullet$\; 
\underline{For $\nu_r \sim 1 \;(1 \le r \le \alpha)$ and $\alpha \to \infty$}\;; 
we simplify the problem by setting $\nu_r=\nu$ for all $1 \le r \le \alpha$, which leads to
\begin{align}
x_r=y_r
&= 
r_{r-2}+\nu y_{r-1} 
\qquad (y_{-1}=0, \quad y_0=1)
\nn\\
& =
\frac{\nu_+^{r+1}-\nu_-^{r+1}}{\nu_+-\nu_-}
\qquad \;(\nu_\pm = \frac{\nu\pm\sqrt{\nu^2+4}}{2})
\nn\\
&\sim \nu^r
\qquad \for \quad 1 \le r \le \alpha.
\end{align} 
Using this, we obtain 
$
\mathcal{L}_j \sim \nu^{2r}
$ for $1 \le j \le m_{\alpha-1}$ and thus 
$
\sum_{j=1}^{m_{\alpha-1}}\mathcal{L}_j\sim y_\alpha^2.  
$
Since we do not have to consider the last component $\{\mathcal{L}_j\}_{j=m_{\alpha-1}+1}^{m_\alpha-1}$ when $1/p_0$ approaches an infinite continued fraction, the spin dc conductivity for $\alpha \to \infty $ is scaled in the high temperature limit as  
\begin{equation}
\mathcal{L}\sim y_\alpha^2.
\end{equation}

\section*{Acknowledgement}
The author is grateful to K. Sakai for valuable discussions and for critical reading of the manuscript.


\begin{appendix}
\section{TS numbers}
Following \cite{TS 72, T 99}, we define series of numbers 
$\{p_r\}_{r=0}^{\alpha+1}, \; \{\nu\}_{r=1}^{\alpha+1}, \; \{m_r\}_{r=0}^{\alpha+1}$ and $\{y_r\}_{r=-1}^\alpha$ starting from the anisotropy parameter $\theta$ as 
\begin{align}
&p_0=\frac{\pi}{\theta}, \quad p_1=1, \quad \nu_r=\left[\frac{p_{r-1}}{p_r}\right], 
\quad p_r=p_{r-2}-\nu_{r-1}p_{r-1}, \nn\\
&p_{\alpha+1}=0, \quad \nu_{\alpha+1}=\infty, \nn\\
&m_0=0, \quad m_r=\sum_{k=1}^r \nu_k, \quad m_{\alpha+1}=\infty, \nn\\
&y_{-1}=0, \quad y_0=1, \quad y_1=\nu_1 \quad \mathrm{and} \quad y_r=y_{r-2}+\nu_r y_{r-1}.  
\label{p, m, y}\end{align}
The TS numbers $\{n_j\}_{j=1}^{m_\alpha}$, the parities $\{v_j\}_{j=1}^{m_\alpha}$ and 
their conjugate numbers $\{q_j\}_{j=1}^{m_\alpha}$ are determined as follows:
\begin{align}
&  n_j=y_{r-1}+(j-m_r)y_r  \qquad (m_r \le  j < m_{r+1}), \nn\\
&  n_{m_\alpha}=y_{\alpha-1},  \nn\\
&  v_{m_1}=-1, \quad v_j=(-1)^{[(n_j-1)/p_0]} \qquad (j \ne m_1)  \nn\\
&  \mathrm{and} \;\; q_j = (-1)^r(p_r-(j-m_r)p_{r+1})  \nn\\
&     \qquad\quad \equiv \frac{1+v_j}{2}p_0-n_j \mod 2p_0   \qquad (m_r \le j < m_{r+1} ). 
\label{TS no.}\end{align}
In this paper, we rewrite $q_j$ as follows by introducing
series of numbers $\{l_r\}_{r=0}^\alpha, \; \{x_r\}_{r=-1}^\alpha$:
\begin{align}
&l_0=0, \quad l_r=\sum_{k=1}^r\nu_{\alpha-(k-1)},
\nn\\
&x_{-1}=0, \quad x_{0}=1, \quad x_1=\nu_\alpha, 
\quad x_r=x_{r-2}+\nu_{\alpha-(r-1)} x_{r-1},
\nn\\
&q_{m_\alpha}=(-1)^\alpha p_\alpha,
\nn\\
& \mathrm{and} \quad  
    q_{m_\alpha-k} = (-1)^{\alpha-(r+1)}(x_{r-1}+(k-l_r)x_r)p_\alpha
    \qquad (l_r < k \le l_{r+1} ).
\label{new q}\end{align}
From this expression we have $q_0=x_\alpha p_\alpha$, which leads to $p_0=x_\alpha p_\alpha$ as we also have $q_0=p_0$ from (\ref{TS no.}). 
On the other hand, continued fraction (\ref{continued fraction}) is rewritten as \\
$1/p_0=x_{\alpha-1}/x_\alpha=x_{\alpha-1}/y_\alpha$. 
Thus we obtain the following relations: 
\begin{equation}
p_0=p_\alpha y_\alpha \quad \textup{and} \quad p_{\alpha}=1/x_{\alpha-1},
\label{p's proof}\end{equation}
where the first relation has been already proved in \cite{Klumper, AS}. 
The modified TS numbers $\{\tilde{n_j}\}_{j=1}^{m_\alpha}$, parities  $\{\tilde{v}_j\}_{j=1}^{m_\alpha}$ and their conjugate numbers $\{\tilde{q_j}\}_{j=1}^{m_\alpha}$ \cite{KSS, AS} are also necessary for this paper:
\begin{align}
&\tilde{n}_j=y_{r-1}+(j-m_r)y_r \qquad (m_r < j \le m_{r+1}), \nn\\ 
&\tilde{v}_j=(-1)^{[(\n_j-1)/p_0]}  \nn\\
&\mathrm{and} \;\; \tilde{q}_j= (-1)^r(p_r-(j-m_r)p_{r+1})  \nn\\
&\qquad\quad \equiv \frac{1+\tilde{v}_j}{2}p_0-\n_j \mod 2p_0   \qquad (m_r < j \le m_{r+1} ).
\label{modified TS no.}\end{align}


\section{Relations (\ref{e's relations}) and (\ref{Tjk simple})}
The rescaled energies $a_j$ and the scattering kernels $T_{j,k}$ satisfy the following relations \cite{TS 72, T 99}:
\begin{align}
&a_j - s_r * ((1 - 2\delta_{m_{r-1},j})a_{j-1} + a_{j+1}) = 0, \nn\\
& \qquad\for \quad m_{r-1} \le j \le m_r-2,    \nn\\
&a_{m_r-1} - (1 - 2\delta_{m_{r-1},m_r-1})s_r * a_{m_r-2}-d_r*a_{m_r-1}
 - s_{r+1}*a_{m_r} = 0, \nn\\
& \qquad \for \quad r<\alpha, \nn\\
&a_{m_\alpha-1}=-a_{m_\alpha}=s_\alpha*a_{m_\alpha-2},
\label{e's complicated}\end{align}
and
\begin{align}
& T_{j,k}-(1-2\delta_{m_{r-1},j})s_r*T_{j-1,k}-s_r*T_{j+1,k} =(-1)^{r+1}(\delta_{j-1,k}+\delta_{j+1,k})s_r, \nn\\
& \qquad\for \quad m_{r-1} \le j \le m_r-2,\quad j\ne m_\alpha-2, \nn\\
& T_{m_r-1,k}-(1-2\delta_{m_{r-1},m_r-1})s_r*T_{m_r-2,k}-d_r*T_{m_r-1,k}-s_{r+1}*T_{m_r,k} \nn\\
& \qquad\quad =(-1)^{r+1}(\delta_{m_r-2,k}s_r+\delta_{m_r-1,k}d_r-\delta_{m_r,k}s_{r+1}), \nn\\
& \qquad \for \quad r<\alpha, \nn\\
&T_{m_\alpha-2,k}-(1-2\delta_{m_{\alpha-1},m_\alpha-2})s_\alpha*T_{m_\alpha-3,k}-s_\alpha*T_{m_\alpha-1,k} \nn\\
& \qquad \quad =(-1)^{\alpha+1}(\delta_{m_\alpha-3,k}+\delta_{m_\alpha-1,k}-\delta_{m_\alpha,k})s_\alpha, \nn\\
& T_{m_\alpha-1,k} =-T_{m_\alpha,k}
=s_\alpha*T_{m_\alpha-2,k}+\varsigma_k\delta_{m_\alpha-2,k}s_\alpha,  \nn\\
\label{Tjk complicated}\end{align}
where
\begin{align}
a_0(\lambda)&=\delta(\lambda), \quad T_{0,k}=0, \nn\\
 s_r(\lambda)&:=\frac{1}{4p_r\ch(\frac{\pi \lambda}{2p_r})} =\int\frac{e^{i\lambda w}}{4\pi \ch(p_r w)}dw, \nn\\
d_r(\lambda)&:=\int\frac{\ch((p_r-p_{r+1})w)e^{i\lambda w}}{4\pi\ch(p_rw)\ch(p_{r+1}w)}dw. 
\label{s, d}\end{align}
We rewrite these relations as  
(\ref{e's relations}) $[\mathbbm{1}-\boldsymbol{S}*]\boldsymbol{a}=0$ and 
(\ref{Tjk simple}) $[\mathbbm{1}-\boldsymbol{S}*]\boldsymbol{T}_k=\boldsymbol{s}_k$ 
by defining matrix $\boldsymbol{S}=(S_{jk})$: 
\begin{align}
&S_{j,j-1}=\begin{matrix*}[l]
(1 - 2\delta_{m_{r-1},j})s_r(\la),  &(m_{r-1} \le j \le m_r-1), 
\end{matrix*} \nn\\
&S_{j,j}=\begin{matrix*}[l]
\delta_{j,m_r-1}(1-\delta_{j,m_\alpha-1})d_r(\la),  
&(m_{r-1} \le j \le m_r-1, \quad j=m_\alpha),
\end{matrix*} \nn\\
&S_{j,j+1}=\begin{matrix*}[l] 
s_r(\la) & ( m_{r-1} \le j \le m_r-1,\quad j\ne m_\alpha-1),  
\end{matrix*} \nn\\
&S_{m_\alpha,m_\alpha-2}=-s_\alpha(\la),  \quad S_{j,k}=0 \quad \textup{for otherwise}, 
\label{matrix S}\end{align}
and vectors $\boldsymbol{s}_k$ which are the columns in the matrix 
$\boldsymbol{s}=(s_{j,k})=(\boldsymbol{s}_1, \cdots, \boldsymbol{s}_k, \cdots, \boldsymbol{s}_{m_\alpha})$:
\begin{align}
&s_{j,k}= \nn\\
&\left\{\begin{matrix*}[l]
(-1)^{r+1}(\delta_{j-1,k}+\delta_{j+1,k})s_r(\la) 
\qquad \for \quad m_{r-1} \le j \le m_r-2, \; j \ne m_\alpha-2 \\
(-1)^{r+1}(\delta_{m_r-2,k}s_r(\la)+\delta_{m_r-1,k}d_r(\la)-\delta_{m_r,k}s_{r+1}(\la))  
\quad \for \quad  j=m_r-1, \quad r<\alpha \\   
(-1)^{\alpha+1}(\delta_{m_\alpha-3,k}+\delta_{m_\alpha-1,k}-\delta_{m_\alpha,k})s_\alpha(\la)  \qquad \for \quad j=m_\alpha-2\\
(-1)^{\delta_{j,m_\alpha}}\varsigma_k\delta_{m_\alpha-2,k}s_\alpha(\la)
\qquad \for \quad j=m_\alpha-1,\; m_\alpha. \\
\end{matrix*}\right. 
\label{higher bare}\end{align}
For example, 
the $\frac{1}{p_0}=\frac{1}{|\nu_1}+\frac{1}{|\nu_2} \;(\nu_1\ge2, \; \nu_2\ge3)$ case is explicitly written as follows:
\begin{align}
& 
\boldsymbol{S}=
\begin{bmatrix}
0& s_1  &           &                  &                  &             &         &         &           &  0         \\   
s_1 & 0 & s_1 \\
      & \ddots & \ddots &\ddots \\
      &           & s_1     &d_1   & s_2 \\  
      &           &           & -s_2             &    0      & s_2  \\
      &           &           &                  &  s_2          & 0    & s_2 \\
      &           &           &                  &                  & \ddots   & \ddots          & \ddots                \\
      &           &           &                  &                  &             & s_2  & 0 & s_2                   \\
      &           &           &                  &                  &            &          & s_2   & 0  & 0          \\
 0   &           &           &                  &                  &            &          &   -s_2  & 0        & 0 
\end{bmatrix}, \\
& \nn\\
&
\boldsymbol{a}=
\begin{bmatrix*}[l]
a_1\\
a_2\\
\vdots \\
a_{\nu_1-1} \\
a_{\nu_1} \\ 
a_{\nu_1+1} \\
\vdots\\
a_{\nu_1+\nu_2-2} \\
a_{\nu_1+\nu_2-1} \\
a_{\nu_1+\nu_2} 
\end{bmatrix*},
\quad
\boldsymbol{T}_k=
\begin{bmatrix*}[l]
T_1,k\\
T_2,k \\
\vdots \\
T_{\nu_1-1,k} \\
T_{\nu_1,k} \\ 
T_{\nu_1+1,k} \\
\vdots\\
T_{\nu_1+\nu_2-2,k} \\
T_{\nu_1+\nu_2-1,k} \\
T_{\nu_1+\nu_2,k} 
\end{bmatrix*},
\quad 
\boldsymbol{s}_k=
\begin{bmatrix}
\delta_{2,k}s_1\\
(\delta_{1,k}+\delta_{3,k})s_1\\
\vdots\\
\delta_{\nu_1-2,k}s_1+\delta_{\nu_1-1,k}d_1-\delta_{\nu_1,k}s_2\\
-(\delta_{\nu_1-1,k}+\delta_{\nu_1+1,k})s_2\\
-(\delta_{\nu_1,k}+\delta_{\nu_1+2,k})s_2\\
\vdots\\
-(\delta_{\nu_1+\nu_2-3,k}+\delta_{\nu_1+\nu_2-1,k}-\delta_{\nu_1+\nu_2,k})s_2\\
\varsigma_k\delta_{\nu_1+\nu_2-2,k}s_2 \\
-\varsigma_k\delta_{\nu_1+\nu_2-2,k}s_2 
\end{bmatrix}.
\end{align}

\section{TBA equations for the spin-1/2 XXZ chain}
Using (\ref{n's relations}), (\ref{e's relations}), (\ref{Tjk simple}), TBA equations (\ref{TBA equations}) are rewritten as follows \cite{TS 72, T 99}:
\begin{align}
\ln\eta_j=&\;(1-2\delta_{m_{r-1},j})s_r*\ln(1+\eta_{j-1})+s_r*\ln(1+\eta_{j+1})
+\delta_{j,1} \beta A s_1  \nn\\
& \for \quad m_{r-1}\le j\le m_r-2, \quad j\ne m_\alpha-2,  \nn \\
\ln \eta_{m_r-1}=&\;(1-2\delta_{m_{r-1},m_r-1})s_r*\ln(1+\eta_{m_r-2})+d_r*\ln(1+\eta_{m_r-1}) \nn\\
&+s_{r+1}*\ln(1+\eta_{m_r})\nn\\
& \for\quad r<\alpha,  \nn\\ 
\ln\eta_{m_\alpha-2}=&\; (1-2\delta_{m_{\alpha-1},m_\alpha-2})s_\alpha*\ln(1+\eta_{m_\alpha-3})+s_\alpha*\ln(1+\eta_{m_\alpha-1})(1+\eta^{-1}_{m_\alpha}), \nn\\
\ln\eta_{m_\alpha-1}-y_\alpha\beta h=&\; y_\alpha\beta h-\ln\eta_{m_\alpha} =s_\alpha*\ln(1+\eta_{m_\alpha-2})   
\label{TBA complicated}
\end{align}
with $\eta_0=0$. $A=-\frac{2\pi J\sin\theta}{\theta}$ is the energy rescaling factor defined by (\ref{rescaling}).
We rewrite these equations as   
(\ref{TBA simple}) 
$\boldsymbol{{\ln\eta}}=(\ln\eta_j)=\beta\g+\boldsymbol{S}*\boldsymbol{\ln(1+\eta)}$
by defining the following vectors:
\begin{equation}
\g=(\mathcal{G}_j),  
\qquad 
\boldsymbol{\ln(1+\eta)}=(\ln(1+\eta)_j), 
\label{NLIEs for 1/2}\end{equation}
where 
\begin{equation}
\mathcal{G}_{j}=\delta_{j,1}As_1(\la)+(\delta_{j,m_\alpha-1}+\delta_{j,m_\alpha})y_\alpha h,
\label{bare energy}\end{equation}
and
\begin{equation}
\ln(1+\eta)_j=
\ln(1+(1-\delta_{j,m_\alpha})\eta_j(\la))(1+\eta_{m_\alpha}(\la)^{-1})^{\delta_{j,m_\alpha-1}}.
\end{equation}
Differentiating equations (\ref{TBA complicated}) with respect to $\beta$ yields
\begin{align}
&\varepsilon_j-(1-2\delta_{m_{r-1},j})s_r*(1-\vartheta_{j-1})\varepsilon_{j-1}-s_r*(1-\vartheta_{j+1})\varepsilon_{j+1}=\delta_{j,1}As_1  \nn\\
& \quad \for\quad m_{r-1}\le j\le m_r-2, \quad j\ne m_\alpha-2,  \nn\\
&\varepsilon_{m_r-1}-(1-2\delta_{m_{r-1},m_r-1})s_r*(1-\vartheta_{m_r-2})\varepsilon_{m_r-2}-d_r*(1-\vartheta_{m_r-1})\varepsilon_{m_r-1} \nn\\
&\qquad -s_{r+1}*(1-\vartheta_{m_r})\varepsilon_{m_r}=0  
\hspace{3cm} \for\quad r<\alpha,  \nonumber\\ 
&\varepsilon_{m_\alpha-2}-(1-2\delta_{m_{\alpha-1},m_\alpha-2})s_\alpha*(1-\vartheta_{m_\alpha-3})\varepsilon_{m_\alpha-3} \nn\\
&\qquad -s_\alpha*(1-\vartheta_{m_\alpha-1}+\vartheta_{m_\alpha})\varepsilon_{m_{\alpha-1}}=0, \nn\\
&\varepsilon_{m_\alpha-1}-y_\alpha h=y_\alpha h-\varepsilon_{m_\alpha} =s_\alpha*(1-\vartheta_{m_\alpha-2})\varepsilon_{m_\alpha-2}  
\label{edr complicated}\end{align}
with $\varepsilon_0=0$. 
We rewrite these relations as   
(\ref{dressed energy}) $[\mathbbm{1}-\boldsymbol{S*(1-\vartheta)}]\boldsymbol{\varepsilon}=\g$
by the vector $\boldsymbol{\varepsilon}=(\varepsilon_j)$ and the matrix $\boldsymbol{S*(1-\vartheta)}=(S*(1-\theta)_{j,k})$: 
\begin{align}
&S*(1-\vartheta)_{j,j-1}=\begin{matrix*}[l]
(1 - 2\delta_{m_{r-1},j})s_r*(1-\vartheta_{j-1}),  &(m_{r-1} \le j \le m_r-1) \\
\end{matrix*}\nn\\
&S*(1-\vartheta)_{j,j}=\begin{matrix*}[l]
\delta_{m_r-1,j}(1-\delta_{m_\alpha-1,j})d_r*(1-\vartheta_j), & (m_{r-1} \le j \le m_r-1, \quad j=m_\alpha) \\
\end{matrix*} \nn\\
&S*(1-\vartheta)_{j,j+1}=\begin{matrix*}[l] 
s_{r+\delta_{m_r-1,j}}*(1-\vartheta_{j+1}+\delta_{m_\alpha-2,j}\vartheta_{m_\alpha}), & ( m_{r-1} \le j \le m_r-1) 
\end{matrix*} \nn\\
&S*(1-\vartheta)_{m_\alpha,m_\alpha-2}=-s_\alpha*(1-\vartheta_{m_\alpha-2}) \nn\\
&S*(1-\vartheta)_{j,k}=0 \quad \textup{for otherwise}. 
\label{S with Fermi}\end{align}

\section{Linear integral equations for $T^\dr_{j,k}$}
Using relations (\ref{Tjk simple}), relations (\ref{dressed Tjk}) are rewritten as
\begin{align}
&T^\dr_{j,k}-(1-2\delta_{m_{r-1},j})s_r*(1-\vartheta_{j-1})T^\dr_{j-1,k}-s_r*(1-\vartheta_{j+1})T^\dr_{j+1,k} \nn\\
&\hspace{2cm}  =(-1)^{r+1}(\delta_{j-1,k}+\delta_{j+1,k})s_r 
\qquad \for \quad m_{r-1} \le j \le m_r-2,\quad j\ne m_\alpha-2, \nn\\
&T^\dr_{m_r-1,k}-(1-2\delta_{m_{r-1},m_r-1})s_r*(1-\vartheta_{m_r-2})T^\dr_{m_r-2,k} 
-d_r*(1-\vartheta_{m_r-1})T^\dr_{m_r-1,k} \nn\\
&\qquad -s_{r+1}*(1-\vartheta_{m_r})T^\dr_{m_r,k} 
=(-1)^{r+1}(\delta_{m_r-2,k}s_r+\delta_{m_r-1,k}d_r-\delta_{m_r,k}s_{r+1})  
\qquad \for \quad r<\alpha, \nn\\
&T^\dr_{m_\alpha-2,k}-(1-2\delta_{m_{\alpha-1},m_\alpha-2})s_\alpha*(1-\vartheta_{m_\alpha-3})T^\dr_{m_\alpha-3,k} \nn\\
&\qquad -s_\alpha*(1-\vartheta_{m_\alpha-1}+\vartheta_{m_\alpha})T^\dr_{m_\alpha-1,k} 
=(-1)^{\alpha+1}(\delta_{m_\alpha-3,k}+\delta_{m_\alpha-1,k}-\delta_{m_\alpha,k})s_\alpha, \nn\\
&T^\dr_{m_\alpha-1,k}=-T^\dr_{m_\alpha,k}=
s_\alpha*(1-\vartheta_{m_\alpha-2})T^\dr_{m_\alpha-2,k}+\varsigma_k\delta_{m_\alpha-2,k}s_\alpha
\label{Tdr complicated}\end{align}
with $T^\dr_{0,k}=0$.
Using vectors (\ref{higher bare}) and matrix (\ref{S with Fermi}), these relations are rewritten as \\ 
(\ref{Tdr linear}) $[\mathbbm{1}-\boldsymbol{S*(1-\vartheta)}]\boldsymbol{T}^\dr_k=\boldsymbol{s}_k$. 
For example, the $\nu_1=\nu_2=3$ case is explicitly written as follows:
\begin{align}
&\boldsymbol{S*(1-\vartheta)} \nn\\
&= 
\begin{bmatrix*}[c]
0                              & s_1*(1-\vartheta_2)   &                             &                                       &                               &0  \\
s_1*(1-\vartheta_1)     &d_1*(1-\vartheta_2)   & s_2*(1-\vartheta_3) \\  
                               & -s_2*(1-\vartheta_2)                                &  0                                     & s_2*(1-\vartheta_4)  \\
                               &                              & s_2*(1-\vartheta_3)  & 0                                     & s_2*(1-\vartheta_5+\vartheta_6)  \\
                               &                              &                              & s_2*(1-\vartheta_4)   & 0                            & 0\\
  0                           &                               &                             & -s_2*(1-\vartheta_4)  & 0                            & 0 
\end{bmatrix*}, \nn \\
& \nn\\
&
\boldsymbol{\varepsilon}=
\begin{bmatrix*}[l]
\varepsilon_1\\
\varepsilon_2\\
\varepsilon_3\\
\varepsilon_4\\
\varepsilon_5\\
\varepsilon_6\\
\end{bmatrix*},
\quad
\boldsymbol{T}^\dr_k=
\begin{bmatrix}
\;T_{1,k}^\dr\;\\
\;T_{2,k}^\dr\;\\
\;T_{3,k}^\dr\;\\
\;T_{4,k}^\dr\;\\
\;T_{5,k}^\dr\;\\
\;T_{6,k}^\dr\;\\
\end{bmatrix},
\quad 
\g=
\begin{bmatrix}
As_1\\
0\\
0\\
0\\
y_\alpha h\\
y_\alpha h\\
\end{bmatrix},\quad 
\boldsymbol{s}_k=
\begin{bmatrix}
\delta_{2,k}s_1\\
\delta_{1,k}s_1+\delta_{2,k}d_1-\delta_{3,k}s_2\\
-(\delta_{2,k}+\delta_{4,k})s_2\\
-(\delta_{3,k}+\delta_{5,k}-\delta_{6,k})s_2\\
\varsigma_k\delta_{4,k}s_2 \\
-\varsigma_k\delta_{4,k}s_2 
\end{bmatrix}.
\end{align}

\section{TBA equations for the spin-$\s/2$ integrable XXZ chain }
If the number of the spin-$\s/2$ is chosen to satisfy the following relation with the string length, the normalizability condition of the string wave function for an infinite chain 
is satisfied---that is, the model is integrable \cite{Kirillov, AS}: 
\begin{equation}
\s+1\in\{\tilde{n}_j\}. 
\end{equation}
In that case, $\s$ is identified as follows by the number $j_\sigma$ with which the number $r_\sigma$ associates uniquely:   
\begin{equation}
\s+1=\tilde{n}_{j_{\s}}, \quad  m_{r_\s-1}< j_\s \le m_{r_\s} .
\label{spin number}\end{equation}
In Ref.\cite{AS}, this condition was expressed as $\s+1=\tilde{n}_{j_{\s}+1}$ and $m_{r_\s-1}\le j_\s < m_{r_\s}$. Following the above new expression (\ref{spin number}), we rewrite 
the TBA equations for the integrable XXZ chain with arbitrary spin-$\s/2$ as 
\begin{align}
&\ln\eta_j^{(j_\s)}=(1-2\delta_{m_{r-1},j})s_r*\ln(1+\eta^{(j_\s)}_{j-1})+s_r*\ln(1+\eta^{(j_\s)}_{j+1}) +(-1)^{r+1}\delta_{j,j_\s-1} A\beta s_r  \nn\\
& \qquad \quad \for \quad m_{r-1}\le j\le m_r-2, \quad j\ne m_\alpha-2,  \nn \\
& \ln \eta^{(j_\s)}_{m_r-1}=(1-2\delta_{m_{r-1},m_r-1})s_r*\ln(1+\eta^{(j_\s)}_{m_r-2})+d_r*\ln(1+\eta^{(j_\s)}_{m_r-1}) \nn\\
&\qquad\qquad\quad +s_{r+1}*\ln(1+\eta^{(j_\s)}_{m_r}) +(-1)^{r+1} \Theta (r_\s+\delta_{m_r,j_\s}-r)A\beta d_r^{(j_\s)} 
\quad \for\quad r<\alpha,  \nn\\ 
&\ln\eta^{(j_\s)}_{m_\alpha-2}=(1-2\delta_{m_{\alpha-1},m_\alpha-2})s_\alpha*\ln(1+\eta^{(j_\s)}_{m_\alpha-3})+s_\alpha*\ln\{1+\eta^{(j_\s)}_{m_\alpha-1}\}\{1+(\eta_{m_\alpha}^{(j_\s)})^{-1}\} \nn\\
&\qquad\qquad\quad +(-1)^{\alpha+1}\delta_{m_\alpha-1,j_\sigma} A\beta s_\alpha, \nn\\
&\ln\eta^{(j_\s)}_{m_\alpha-1}-y_\alpha\beta h
=y_\alpha\beta h-\ln\eta^{(j_\s)}_{m_\alpha} 
=s_\alpha*\ln(1+\eta^{(j_\s)}_{m_\alpha-2}) +(-1)^{\alpha+1}\delta_{m_\alpha,j_\s}A\beta s_\alpha,  
\label{higher TBA}\end{align}
where   
\begin{align}
&\eta_0^{(j_\s)}=0,  \nn\\
&d_r^{(j_\s)}(\lambda):=\int_{-\infty}^\infty\frac{\ch(\tilde{q}_{j_\s}w)e^{i\lambda w}}{4\pi\ch(p_{r+1}w)\ch(p_r w)}dw, \nn\\
&\Theta (r):=\left\{\begin{array}{l}1\quad (r \ge 0) \\0 \quad (r <0). \end{array}\right.
\end{align} 
We further rewrite these equations as (\ref{higher spin TBA simple})
$\boldsymbol{\ln\eta}\boldsymbol{\eta}^{(j_\s)}=(\ln\eta_j^{(j_\s)})=\beta\g^{(j_\s)}+\boldsymbol{S}*\boldsymbol{\ln(1+\eta)}^{(j_\s)}$
by defining the following vectors:
\begin{equation}
\g^{(j_\s)} =(\mathcal{G}_j^{(j_\s)}), 
\qquad 
\boldsymbol{\ln(1+\eta)}^{(j_\s)}=(\ln(1+\eta)_j^{(j_\s)}),  
\label{NLIEs for s/2}\end{equation}
where 
\begin{align}
&\mathcal{G}_j^{(j_\s)}=
(-1)^{r+1}A
\{(\delta_{j,j_\s-1}-\delta_{m_\alpha,j}\delta_{m_\alpha,j_\s}) s_r   
+\delta_{m_r-1,j}\Theta (r_\s-r)d_r^{(j_\s)}\} \nn\\
&\qquad\quad +(\delta_{m_\alpha-1,j}+\delta_{m_\alpha,j})y_\alpha h   
\qquad\quad \for \quad m_{r-1} \le j \le m_r-1,\quad  j=m_\alpha,
\end{align}
and
\begin{equation}
\ln(1+\eta)_j^{(j_\s)}=\ln\{1+(1-\delta_{m_\alpha,j})\eta_j^{(j_\s)}(\la)\}\{1+\eta_{m_\alpha}^{(j_\s)}(\la)^{-1}\}^{\delta_{m_\alpha-1,j}}.  
\end{equation}
Differentiating equations (\ref{higher TBA}) with respect to $\beta$, we obtain
\begin{align}
\varepsilon_j^{(j_\s)}=&\;(1-2\delta_{m_{r-1},j})s_r*(1-\vartheta_{j-1}^{(j_\s)})\varepsilon_{j-1}^{(j_\s)}+s_r*(1-\vartheta_{j+1}^{(j_\s)})\varepsilon_{j+1}^{(j_\s)} +(-1)^{r+1}\delta_{j,j_\s-1}A s_r  \nn\\
& \for\quad m_{r-1}\le j\le m_r-2, \quad j\ne m_\alpha-2,  \nn\\
\varepsilon_{m_r-1}^{(j_\s)}=&\;(1-2\delta_{m_{r-1},m_r-1})s_r*(1-\vartheta^{(j_\s)}_{m_r-2})\varepsilon_{m_r-2}^{(j_\s)}+d_r*(1-\vartheta^{(j_\s)}_{m_r-1})\varepsilon_{m_r-1}^{(j_\s)} \nn\\
&+s_{r+1}*(1-\vartheta^{(j_\s)}_{m_r})\varepsilon_{m_r}^{(j_\s)} +(-1)^{r+1} \Theta (r_\sigma+\delta_{m_r,j_\s}-r)d_r^{(j_\s)} \nn\\
&\for\quad r<\alpha,  \nonumber\\ 
\varepsilon_{m_\alpha-2}^{(j_\s)}=&\;(1-2\delta_{m_{\alpha-1},m_\alpha-2})s_\alpha*(1-\vartheta_{m_\alpha-3}^{(j_\s)})\varepsilon^{(j_\s)}_{m_\alpha-3}+s_\alpha*(1-\vartheta^{(j_\s)}_{m_\alpha-1}+\vartheta^{(j_\s)}_{m_\alpha})\varepsilon^{(j_\s)}_{m_{\alpha-1}} \nn\\
&+(-1)^{\alpha+1}\delta_{m_\alpha-1,j_\s} s_\alpha, \nn\\
\varepsilon_{m_\alpha-1}^{(j_\s)}=&-\varepsilon_{m_\alpha}^{(j_\s)} 
=s_\alpha*(1-\vartheta^{(j_\s)}_{m_\alpha-2})\varepsilon_{m_\alpha-2}^{(j_\s)} +(-1)^{\alpha+1}\delta_{m_\alpha,j_\s}s_\alpha  
\end{align}
with $\varepsilon_0^{(j_\s)}=0$. 
We rewrite these relations as
\begin{equation}
[\mathbbm{1}-\boldsymbol{S*(1-\vartheta)}^{(j_\s)}]\boldsymbol{\varepsilon}^{(j_\s)}=\g^{(j_\s)}, 
\end{equation}
where the matrix $\boldsymbol{S*(1-\vartheta)}^{(j_\s)}=(S*(1-\vartheta^{(j_\s)})_{j,k})$ is defined as    
\begin{align}
&S*(1-\vartheta^{(j_\s)})_{j,j-1}=\begin{matrix*}[l]
(1 - 2\delta_{m_{r-1},j})s_r*(1-\vartheta^{(j_\s)}_{j-1}),  &(m_{r-1} \le j \le m_r-1) \\
\end{matrix*}\nn\\
&S*(1-\vartheta^{(j_\s)})_{j,j}=\begin{matrix*}[l]
\delta_{m_r-1,j}(1-\delta_{m_\alpha-1,j})d_r*(1-\vartheta^{(j_\s)}_j), & (m_{r-1} \le j \le m_r-1, \quad j=m_\alpha) \\
\end{matrix*} \nn\\
&S*(1-\vartheta^{(j_\s)})_{j,j+1}=\begin{matrix*}[l] 
s_{r+\delta_{m_r-1,j}}*(1-\vartheta^{(j_\s)}_{j+1}+\delta_{m_\alpha-2,j}\vartheta^{(j_\s)}_{m_\alpha}), & ( m_{r-1} \le j \le m_r-1) 
\end{matrix*} \nn\\
&S*(1-\vartheta^{(j_\s)})_{m_\alpha,m_\alpha-2}=-s_\alpha*(1-\vartheta^{(j_\s)}_{m_\alpha-2}), \nn\\
&S*(1-\vartheta^{(j_\s)})_{j,k}=0 \quad \textup{for otherwise}. 
\label{higher S with Fermi}\end{align}
Up to the first order of $\beta A$, we obtained the high temperature expansions of the solutions to equations (\ref{higher TBA})  as   
\begin{align}
\eta^{(j_\s)}_j(\la)
& =
\left(\frac{\n_{j+1}}{y_r}\right)^2
\left[
1+\frac{\beta A}{y_r\n_{j+1}}\sum_{s=1}^{\s}\frac{s(\s+1-s)}{\s+1}
\Delta a^{(j_\s)}_{j,s} (\la)
\right]-1
\nn\\ &\quad
+O(\beta y_\alpha h)+O\left((\beta A)^2, \; \beta^2 y_\alpha hA\right)  
\nn\\ 
\for \quad & m_r\le j < m_{r+1}, \quad j \ne m_\alpha-1, \; m_\alpha,  
\nn\\
\eta^{(j_\s)}_{m_\alpha-1}(\la)
&=
\left(\eta^{(j_\s)}_{m_\alpha}(\la)\right)^{-1} 
\nn\\ &=
\frac{y_\alpha}{y_{\alpha-1}}\left[1+\frac{\beta A}{y_{\alpha-1}}\sum_{s=1}^{\s}\frac{s(\s+1-s)}{\s+1} 
a^{(j_\s)}_{m_\alpha-1,s} (\la)\right]-1 
\nn\\ &\quad
+O(\beta y_\alpha h)+O\left((\beta A)^2, \; \beta^2 y_\alpha hA\right).
\label{limit of eta}\end{align}
We performed these expansions in the same way as we did in Section 5 of Ref.\cite{AS}. Note that the $O(\beta y_\alpha h)$ terms in these expansions are independent of the spectral parameter $\la$, from which it follows that these terms do not enter into the expression (\ref{L at inifinite temp}) for $\mathcal{L}$ after taking the derivatives with respect to $A\beta$ or $\la$.
The functions $\Delta a^{(j_\s)}_{j,s} (\la)$ and $a^{(j_\s)}_{j,s} (\la)$ are defined by (\ref{Intro's a}) and (\ref{higher a}) respectively.

\end{appendix}


\end{document}